\documentclass[twocolumn]{aastex631}

\usepackage{amsmath}	
\usepackage{amssymb}	

\usepackage{color}
\usepackage{xcolor}

\submitjournal{ApJS}

\shorttitle{A dynamical survey of the trans-Neptunian region I.}
\shortauthors{Forg\'acs-Dajka et al.}

\graphicspath{{./}{figures/}}

\begin{document}

\title{A dynamical survey of the trans-Neptunian region I.: Mean-motion resonances with Neptune}

\correspondingauthor{Cs. Kiss}
\email{kiss.csaba@csfk.org}

\author[0000-0002-5735-6273]{E. Forgács-Dajka} 
\affiliation{Department of Astronomy, Institute of Geography and Earth Sciences, E\"otv\"os Lor\'and University,\\
H-1117 Budapest, P\'azm\'any P\'eter s\'et\'any 1/A, Hungary}
\affiliation{Centre for Astrophysics and Space Science, E\"otv\"os Lor\'and University,\\
H-1117 Budapest, P\'azm\'any P\'eter s\'et\'any 1/A, Hungary}
\affiliation{ELKH-SZTE Stellar Astrophysics Research Group, H-6500 Baja, Szegedi út, Kt. 
766, Hungary}
\affiliation{Wigner Research Centre for Physics, P.O. Box 49, Budapest H-1525, Hungary}

\author[0000-0002-4491-2824]{E. K\H ov\'ari} 
\affiliation{Department of Astronomy, Institute of Geography and Earth Sciences, E\"otv\"os Lor\'and University,\\
H-1117 Budapest, P\'azm\'any P\'eter s\'et\'any 1/A, Hungary}
\affiliation{Centre for Astrophysics and Space Science, E\"otv\"os Lor\'and University,\\
H-1117 Budapest, P\'azm\'any P\'eter s\'et\'any 1/A, Hungary}
\affiliation{Wigner Research Centre for Physics, P.O. Box 49, Budapest H-1525, Hungary}

\author[0000-0002-0697-6050]{T. Kov\'acs}
\affiliation{ELKH--ELTE Extragalctic Astrophysics Research Group, E\"otv\"os Lor\'and University,\\
H-1117 Budapest, P\'azm\'any P\'eter s\'et\'any 1/A, Hungary}

\author[0000-0002-8722-6875]{Cs. Kiss}
\affiliation{Konkoly Observatory, Research Centre for Astronomy and Earth Sciences, Konkoly Thege 15-17, H-1121, Budapest, Hungary}
\affiliation{CSFK, MTA Centre of Excellence, Budapest, Konkoly Thege Miklós út 15-17., H-1121, Hungary}
\affiliation{ELTE E\"otv\"os Lor\'and University, Institute of Physics, Budapest, Hungary}

\author[0000-0003-1216-913X]{Zs. S\'andor}
\affiliation{Department of Astronomy, Institute of Geography and Earth Sciences, E\"otv\"os Lor\'and University,\\
H-1117 Budapest, P\'azm\'any P\'eter s\'et\'any 1/A, Hungary}
\affiliation{Centre for Astrophysics and Space Science, E\"otv\"os Lor\'and University,\\
H-1117 Budapest, P\'azm\'any P\'eter s\'et\'any 1/A, Hungary}
\affiliation{Konkoly Observatory, Research Centre for Astronomy and Earth Sciences, Konkoly Thege 15-17, H-1121, Budapest, Hungary}

\begin{abstract}
In this paper, we present a large-scale dynamical survey of the trans-Neptunian region, with particular attention to mean-motion resonances (MMRs). We study a set of 4121 trans-Neptunian objects (TNOs), a sample far larger than in previous works. We perform direct long-term numerical integrations that enable us to examine the overall dynamics of the individual TNOs as well as to identify all MMRs. For the latter purpose, we apply the own-developed FAIR method that allows the semi-automatic identification of even very high-order MMRs. Apart from searching for the more frequent eccentricity-type resonances that previous studies concentrated on, we set our method to allow the identification of inclination-type MMRs, too. Furthermore, we distinguish between TNOs that are locked in the given MMR throughout the whole integration time span ($10^8$\,years) and those that are only temporarily captured in resonances.  For a more detailed dynamical analysis of the trans-Neptunian space, we also construct dynamical maps using test particles. Observing the fine structure of the $ 34-80 $~AU region underlines the stabilizing role of the MMRs, with the regular regions coinciding with the position of the real TNOs.
\end{abstract}

\keywords{1396:Resonant Kuiper belt objects, 1082:N-body simulations}

\section{Introduction \label{sec:intro}}

The outer region of the Solar System - the trans-Neptunian space, or habitually referred to as the Kuiper belt - contains a large population of small bodies which are generally regarded as the remnants of the planet-forming population of planetesimals  \citep{2008ssbn.book..335B,2014AJ....147....2S,2014Natur.507..471T,2016AJ....151...22B}. These trans-Neptunian objects (TNOs) are divided into subclasses based on their dynamical behaviour \citep{2008ssbn.book...43G}: a large population of small bodies with small eccentricities and small-to-moderate inclinations form the \emph{classical Kuiper belt}, and objects on eccentric and inclined orbits belong to the \emph{scattered disk}, including the \emph{detached objects} with large perihelion distances. Objects in mean-motion resonance with Neptune (a.k.a. the resonant TNOs) are considered as a distinct dynamical class as their motion is chiefly governed by their actual resonance. 

The dynamics of the Kuiper belt are determined by the current environment around the Sun: at its outer perimeter, the Kuiper belt ends at $a$\,$=$\,2000\,AU, where the inner Oort cloud begins. Beyond that, the Galactic environment starts to dominate the dynamics by being able to alter the perihelion distance $q$ and the inclination $I$ on an orbital timescale. The outer Oort cloud begins at $a=$\,10000\,AU where the galactic tide makes its structure spherical. We note though that it is possible that in the early Solar System, the above boundaries were located at different heliocentric distances, due to the different initial conditions. In our study, we focus on the distance range $ 30.1 < a < 2000 $ AU, which is the same as that of the database of the TNOs, as explained below.

The orbital motion of the TNOs encompasses a wide dynamical spectrum. Their dynamics are governed by various effects such as close encounters \citep{2004Natur.427..518F}, secular effects of giant planets, mean-motion resonances with Neptune \citep{2019GSL.....6...12M}, chaotic scattering due to the overlapping of adjacent MMRs \citep{2008ssbn.book..275M}, effects from the Galactic tide \citep{1986Icar...65...13H}, and incidental effects due to nearby passing stars \citep{2003EM&P...92....1M}. If we consider comets, too, one can see that in contrast to its apparent dynamical stability, the Kuiper belt is permanently eroding. Cometary nuclei are drifting towards the inner Solar System due to gravitational perturbations of planets and dwarf planets (acting mainly on secular timescales) \citep[see e.g. the review of][]{1998CeMDA..72..129M}.

It is important to note that the above-cited studies are, irrespective of their numerical or theoretical character, all based on the observations of distant minor bodies; therefore, one encounters different patterns in the distribution of the orbital elements of the observed bodies. These patterns might be indicators of a novel or so-far hidden physical processes in the outer Solar System or are generated by observational biases. The detailed treatment of these patterns is summarized in the review paper by \cite{2020tnss.book...61K}. In this study, the authors demonstrate that the reason for the observational biases is due to the process of how TNOs are discovered. Moreover, the observational biases can be reduced or eliminated by a careful design of the observational surveys and various physical phenomena can also lead to distortions in the orbital elements' distributions.

Earlier works dealing with the stability of the outer Solar System have mainly focused on the transport mechanisms of comets \citep{2001Icar..152....4R}. However, several other works have been published that aim at investigating the dynamical behaviour of the whole population as well as its formation using both theoretical and numerical approaches \citep[see e.g. the review of][]{2020CeMDA.132...12S}.

In this paper, we study the dynamics of all known TNOs with a special emphasis on objects involved in MMRs with Neptune. Resonances are particularly interesting as they may provide protection against perturbations and allow large-eccentricity orbits to survive for the age of the Solar System. Moreover, temporary trapping of objects near the border of the resonance is possible, and likewise, nearly resonant objects can escape into a dynamical domain where perturbations may drive them out of the Kuiper belt. Resonant TNOs are also important diagnostic tools for the long-term evolution of the Solar system, in particular for the planet migration era. The majority of the present resonant populations were likely emplaced during the planet migration process \citep{Gladman2012}. Resonant TNOs may be connected to the scattered disk \citep{Gomes2008} population, and recent studies \citep{Yu2018} indicate that a significant fraction of all scattered objects are transiently stuck in MMRs, suggesting that these objects originate from the same single population. 
The dynamical behaviour combined with physical characteristics provides a unique tool to test the formation conditions and the subsequent evolution of these objects \citep{Lacerda2014,FT20}. 

Our paper is organised as follows: we summarize in Section~\ref{sec:data_and_methods} the source of our data and we introduce the methods used. We briefly review our recently developed FAIR method which enables the fast and semi-automatic identification of resonant objects and presents the chaos indicators applied in the construction of dynamical maps. In Section~\ref{sec:the_res_struct} we review the resonant structure of the trans-Neptunian region. We present a statistical overview of the identified resonant objects; we distinguish between the so-called eccentricity- and inclination-type MMRs and give the general abundance of these MMRs. We also differentiate between those TNOs whose critical argument librates during the whole numerical integration and those with only temporarily librating critical arguments. We also give examples of dynamically interesting and peculiar objects. Using a huge number of test particles, we reveal the dynamical structure of the trans-Neptunian region by compiling very detailed dynamical maps. In Section~\ref{sec:summa}, we summarize our results.

\section{Data and methods}
\label{sec:data_and_methods}

\subsection{Data source and the numerical integrator}

Due to the extensive observational surveys of the faraway realms of the Solar System, the number of TNOs increased considerably in recent years, yet their identification is more difficult than that of the asteroids in the main belt, as a consequence of their more distant orbits. Traditionally, the trans-Neptunian region is divided into the Kuiper belt, the scattered disc, and the Oort cloud, which classification is based on the growing distance of the objects in question from the Sun. In this work, we do not make such distinctions but concentrate instead on all the objects classified as TNOs, (i.e., small bodies with a semi-major axis $a>30.1$ AU). 

We used NASA's JPL Horizons database\footnote{https://ssd.jpl.nasa.gov/horizons/} to identify and collect all the TNOs (satisfying the above criterion), and also to obtain the coordinates and velocities of these objects. Our sample includes 4121 objects in total. 

We note here that the data of the TNOs have uncertainties, and their magnitudes depend on the number and quality of the observations. To take this into account, JPL provides a so-called Orbital Condition Code (OCC) for each object. The Orbit Condition Code is an integer between 0 and 9 indicating on a logarithmic scale how well an object's orbit is known. 0 represents the best and 9 the poorest orbit quality. TNOs with an OCC $>$ 5 are generally considered ''lost'' for the purposes of locating them again in the sky at future apparitions. An alternative for JPL's OCC is the Minor Planet Center's (MPC's) ''U'' parameter\footnote{ http://www.minorplanetcenter.net/iau/info/UValue.html}.

In a previous study \citep{2022A&A...657A.135F} we investigated the impact of the refinement of orbital elements from a statistical point of view. We downloaded the data of the Hungaria asteroid family available both in 2019 and 2020 and carried out a statistical comparison of the data with regard to MMRs. According to our investigation, the given MMRs changed neither quantitatively, nor qualitatively. In the case of individual bodies, an object might have been categorized as resonant based on its 2019 data, but non-resonant based on its 2020 data or vice versa. However, the statistical properties of the entire sample remained unaltered i.e. the fraction of resonant objects was the same, regardless of the year of origin of the dataset.
These observations should be kept in mind in the case of the present study, too, although we did not repeat the above comparative study for the TNOs.

In addition to these observational uncertainties, selection effects should also be addressed when studying such distant objects. Obviously, the probability to discover a faraway body is highly dependent on its distance, but also on the small body's physical properties (size, albedo, etc.).
However, despite the limitations of detection and the imperfection of the database, we believe that our study brings us closer to a deeper understanding of the dynamics of the trans-Neptunian space.

To identify all the MMRs of the sample as well as to study the long-term dynamics of the TNOs, we carried out direct numerical integrations. We took into consideration - apart from the given TNO and the Sun - the four giant planets (Jupiter, Saturn, Uranus, and Neptune), and adapted a barycentric coordinate system. We then integrated the equations of motion by using a self-developed adaptive step size Runge--Kutta--Nyström 6/7 $N$-body integrator \citep{1978CeMec..18..223D}, in which the tolerance was set to $10^{-14}$ to ensure the accuracy during the long integration time. The barycentric orbital elements that we used in our work have been calculated from the barycentric coordinates and velocities obtained by the numerical integrations.

\subsection{The FAIR method and its application}

The FAIR method is a recent and efficient method to quickly identify bodies involved in MMRs and is applicable both in the case of the Solar System and exoplanetary systems \citep{2018MNRAS.477.3383F}.

In the case of an MMR, the ratio of the mean motions $ n $ and $ n'$ of the bodies involved can be expressed as the ratio of small positive integers $ p $ and $ q $: $n/n^\prime \approx (p+q)/p$. We note here that the primed quantities refer to the perturbing body, in our case Neptune. However, the above formula gives only a necessary but not sufficient condition. To obtain a sufficient condition, the libration of the critical argument corresponding to the MMR under study should also be fulfilled. Thus the identification of higher-order MMRs could be cumbersome if it is based solely on the calculation of the ratio of the semi-major axes and mean motions. To decide whether the bodies are in an MMR in this case, preliminary guesses are to be made. The FAIR method, however, offers an elegant solution to the identification of any MMR without the a priori knowledge of the critical argument.

The critical argument corresponding to an outer eccentricity-type MMR has coefficients in the series expansion of the perturbing function containing $q$th-order terms in the eccentricities $e$ or $e^{\prime}$. When only terms in the form $e^{|q|}$ are present, the critical argument takes the form
\begin{equation}
    \theta_{1}^{(e)} = (p+q)\lambda - p\lambda^{\prime} - q\varpi.
    \label{eq:crit_arg_e}
\end{equation}
When only terms in form $e^{\prime |q|}$ appear, the corresponding critical argument is
\begin{equation}
    \theta_{2}^{(e^{\prime})} = (p+q)\lambda - p\lambda^{\prime} - q\varpi^{\prime}.
    \label{eq:crit_arg_eprime}
\end{equation}
In Equations \eqref{eq:crit_arg_e}-\eqref{eq:crit_arg_eprime}, $ \lambda $ and $ \lambda' $ denote the mean longitudes of the bodies, and $ \varpi $ and $ \varpi' $ stand for the longitudes of perihelion.

In addition to the more frequently examined eccentricity-type MMRs, one could also take into consideration the so-called inclination-type MMRs. They are related to the precession of the ascending nodes $ \Omega $ and $ \Omega' $ that happens in the case of an oblate central body (e.g. in the satellite system of Neptune). On the other hand, when a forced convergent migration takes place among giant planets, first, they are captured into eccentricity-type MMRs, which can then be followed by capture into an inclination-type MMR - due to the excitation of their mutual inclination \citep{2003ApJ...597..566T,2009MNRAS.400.1373L}. Since the orbital architecture of TNOs shows clear signs of formation via the outward migration of Neptune, several inclination-type MMRs with Neptune might supposedly be present among them. In the series expansion of the perturbing function, the critical arguments of pure outer inclination-type MMRs are written in the forms
\begin{equation}
    \theta_{1}^{(I)} = (p+q)\lambda - p\lambda^{\prime} - q\Omega
    \label{eq:crit_arg_inc}
\end{equation}
and
\begin{equation}
    \theta_{2}^{(I^{\prime})} = (p+q)\lambda - p\lambda^{\prime} - q\Omega^{\prime}
    \label{eq:crit_arg_incprime}
\end{equation}
with the same denotations as above.

We note here that in this paper we do not consider the mixed eccentricity- and inclination-type MMRs, whose critical arguments appear in the cases of higher-order expansions of the perturbing function.

Restricting our investigation to outer MMRs (which is the case of the TNOs), we briefly review the FAIR method that is applicable both in the cases of non-mixed pure eccentricity- and inclination-type resonances.

If one displays the difference between the mean longitudes of the bodies involved in an MMR, i.e., $\lambda - \lambda^{\prime}$ as either the function of the mean anomaly $ M $ or that of $ M' $, several stripes appear whenever an eccentricity-type MMR is found. By counting the intersections of these stripes with the horizontal and vertical axes the degree ($p$) and the order ($q$) of the MMR can easily be identified (see Table 1 in \cite{2018MNRAS.477.3383F}), thus the critical argument can be constructed, too. For instance, if there are $q$ intersections with the horizontal and $p+q$ intersections with the vertical axis, the corresponding critical argument is given by Equation \eqref{eq:crit_arg_eprime}.

The above procedure can also be applied to inclination-type MMRs. In that case, $\lambda - \lambda^{\prime}$ should be plotted as the function of $M+\omega$ or $M^{\prime}+\omega^{\prime}$. Again, by counting the intersections of the stripes with the horizontal and vertical axes, the corresponding critical argument can easily be written. 

A detailed description of the above procedure can be found in \cite{2018MNRAS.477.3383F}. We emphasize here that this method is applicable to a large set of objects using a quasi-automatic approach (see our recent work \citep{2022A&A...657A.135F} where we classified approximately 25000 Hungaria asteroids as either resonant or non-resonant bodies, along with the degree and order of the resonance in the former case).

\subsection{Chaos indicators based on the variation of action-like variables} 

In the last decades, there has been a growing interest in the development and application of various chaos detection methods when studying the dynamical behaviour of dynamical systems. The most classic method is the calculation of the largest Lyapunov Characteristic Exponent (LCE), in which the time evolution of the tangent vector to the instantaneous space vector is calculated \citep{1980Mecc...15....9B}. One limitation of the calculation of the LCE is, however, that it can only be obtained as a limit; therefore, one cannot be sure whether the trajectory under study is regular or just weakly chaotic. To decide about a system's dynamics in a reasonably short time, several new methods have been developed. By using these methods, one can detect the regular or irregular nature of individual trajectories, as well as separate regions of regular and chaotic motions. 

Such a method is the Fast Lyapunov Indicator (FLI) \citep[see][]{1997CeMDA..67...41F}. This allows to distinguish between chaotic and regular motion; and moreover, between quasi-periodic and resonant orbits, too. It is worth mentioning that the dynamics of 716 main belt asteroids - of orbits in between the $3:1$ and $5:2$ Kirkwood gaps - have been studied by using this method. Another frequently adopted method is the Mean Exponential Growth of Nearby Orbits (MEGNO) \citep{2000A&AS..147..205C}, which is efficient both to study individual orbits and to separate the regions of regular and chaotic motion. With a relatively modest computational potential, even the fine structure of a chosen parameter space can be determined. Finally, the method of the Relative Lyapunov Indicator (RLI) has to be mentioned whose calculation is based on the short-time approximation of the LCE of two neighbouring trajectories \citep{2004CeMDA..90..127S}. This approach can also be employed to characterize the dynamical structure of a given parameter plane.

We note that all of these methods are based on the time evolution of the infinitesimally small tangent vector to the trajectory in the phase space, which is provided by the variational equations that should be solved together with the equations of motion. The above chaos detection methods are referred to as variational indicators.

In the cases of dynamical systems with many degrees of freedom, the calculation of the variational indicators can, however, be quite complicated. Therefore, in such cases, it is recommended to use the maximum variation of either the eccentricities or the semi-major axes to reveal the dynamical characteristics and structure of the region under study. Such an indicator can be calculated by fixing the integration time and considering the minimum and the maximum values of either $e$ or $a$, thus both $\Delta e:= \mathrm{max}(e) - \mathrm{min}(e)$ and $\Delta a:= \mathrm{max}(a) - \mathrm{min}(a)$ can easily be calculated, for instance. Plotting the values of either $\Delta e$ or $\Delta a$ on the parameter planes $(a,e)$ or $(a,I)$, i.e., producing the dynamical maps of a system, the overall dynamical structure, along with resonances and their separatrices as well as chaotic layers attached to them can be visualized. A recent application of this simple method is provided in \cite{2022A&A...657A.135F} where the authors investigated the resonant dynamics of the Hungaria asteroids by mapping the neighbourhood of the most populated MMRs. We have also made a benchmark calculation by using the MEGNO indicator and obtained the same stability map as in the case of $\Delta a$ and $\Delta e.$ Therefore, we adopt this latter technique to explore the dynamical properties of the trans-Neptunian region.

\section{The resonant structure of the trans-Neptunian space}
\label{sec:the_res_struct}

In the following subsections, we review the dynamical structure of the trans-Neptunian space, with a particular interest in the resonances. The FAIR method, presented earlier, enables the identification of the specific MMRs in between Neptune and the given TNO, without any a priori information. In this study, the eccentricity- and inclination-type resonances will be examined in detail. Table~\ref{tab:nameofmmr} gives an excerpt of our large-scale survey about the identification of MMRs (by using the FAIR method) and libration types (direct numerical integration) of the catalogued TNOs. The last column of the table contains the measure of the orbit determination's uncertainty, the OCC value. As mentioned earlier, an OCC $> 5$ implies significant uncertainty, thus in the corresponding cases, the resonant properties of the TNO should be treated accordingly. Nevertheless, we did not exclude these cases from our study, for we believe that for statistical purposes they are usable.

In addition to the individual TNOs, we construct dynamical maps, too, which can be used to distinguish between the stable/unstable regions, reveals the structure of mean-motion resonances, illustrate the secular effects, and show the chaotic regions resulting from the overlap of resonances.

\begin{deluxetable}{lcccc}
\tablecaption{Summary of the resonant properties of trans-Neptunian objects\label{tab:nameofmmr}. $1^\mathrm{st}$ column: provisional designation; $ 2^\mathrm{nd} $ column: ratio of the mean-motion resonance; $ 3^\mathrm{rd} $ column: type of the MMR (e: eccentricity-type, I: inclination-type, LT: long-term, ST: short-term MMR); $ 4^\mathrm{th} $ column: amplitude of libration in degrees; $ 5^\mathrm{th} $ column: Orbit Condition Code}. The entire table consisting of 1359 resonant TNOs is available in a machine-readable format.
\tablewidth{0pt}
\tablehead{
\colhead{\shortstack{Desig.}} & \colhead{\shortstack{MMR}} & \colhead{\shortstack{Type}} & \colhead{\shortstack{Ampl. [deg]}} & \colhead{\shortstack{OCC}} }
\startdata
	 2012\,HH$_{2}$       & 5:4          &  e:LT/I:LT       &  53.60/61.78      &  2 \\    
     2015\,SV$_{20}$      & 5:4          &  e:LT/I:ST       &  60.38/-          &  2 \\        
     1997\,QJ$_{4}$       & 3:2          &  e:LT/I:LT       &  93.59/134.51     &  2 \\           
     2001\,KX$_{76}$      & 3:2          &  e:LT/I:LT       &  65.36/104.76     &  3 \\    
     2015\,RB$_{278}$     & 4:1          &  e:LT/I:LT       &  52.67/105.71     &  4 \\   
     2015\,GP$_{54}$      & 3:2          &  e:LT			&  123.38           &  4 \\ 
     2016\,PA$_{101}$     & 14:9         &  e:ST            &  -                &  6 \\         
     2014\,QH$_{563}$     & 13:9         &  I:ST            &  -                &  4       
\enddata
\end{deluxetable}

\subsection{Distribution of the orbital elements of TNOs}
\begin{figure}
\centering
\includegraphics[width=0.95\linewidth]{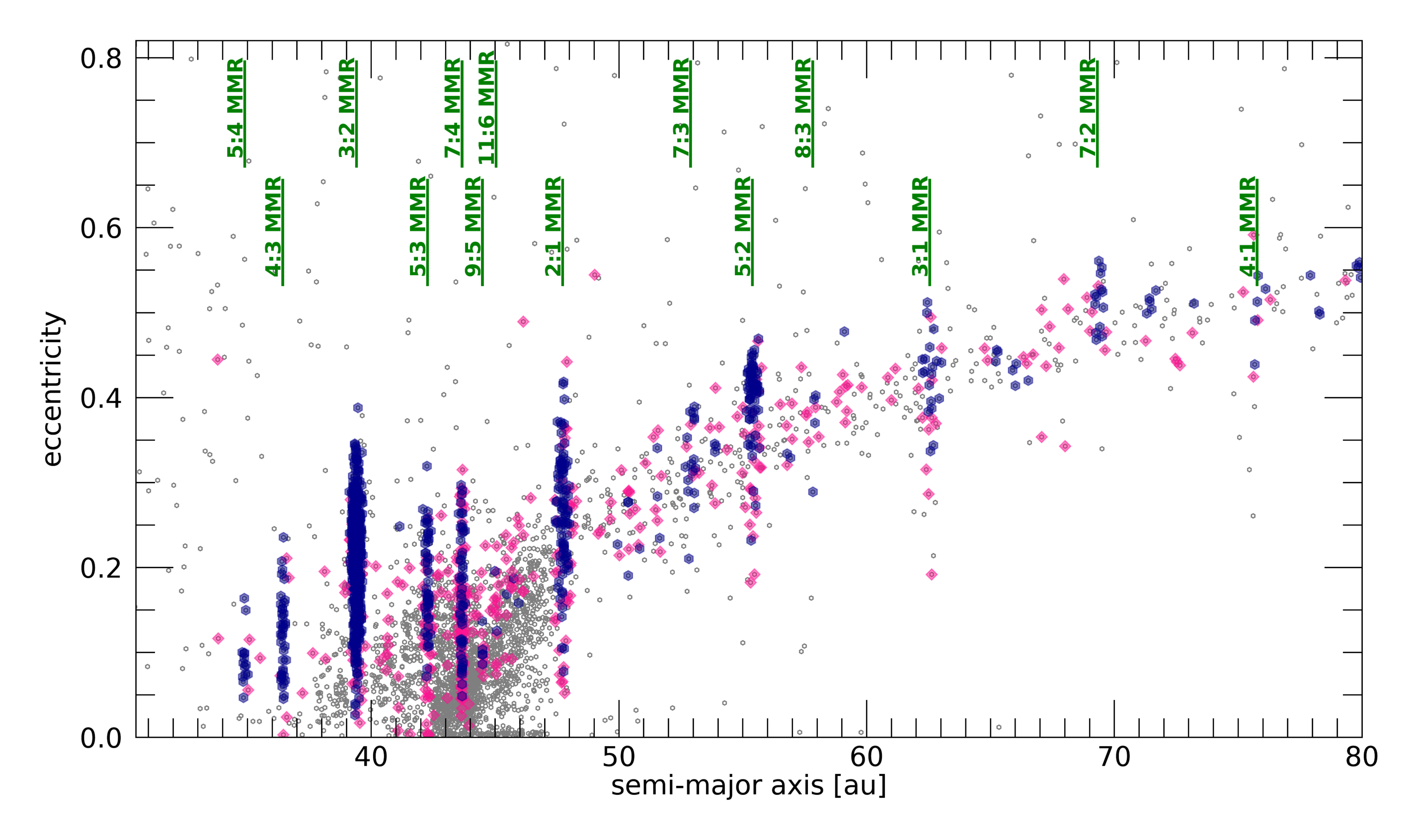}
\caption{\label{fig:tno-all-and-ecc-mmr-sma-ecc}
Distribution of TNOs in the $(a,e)$ plane. Grey dots denote non-resonant TNOs, coloured ones stand for eccentricity-type resonant TNOs: blue dots indicate TNOs that are engaged in the given MMR throughout the whole $10^8$-year-long time span of the numerical integration, pink dots mark TNOs that are captured temporarily in the given MMR. Vertical green lines highlight the location of MMRs with more than 10 members.
}
\end{figure}

\begin{figure}
\centering
\includegraphics[width=0.95\linewidth]{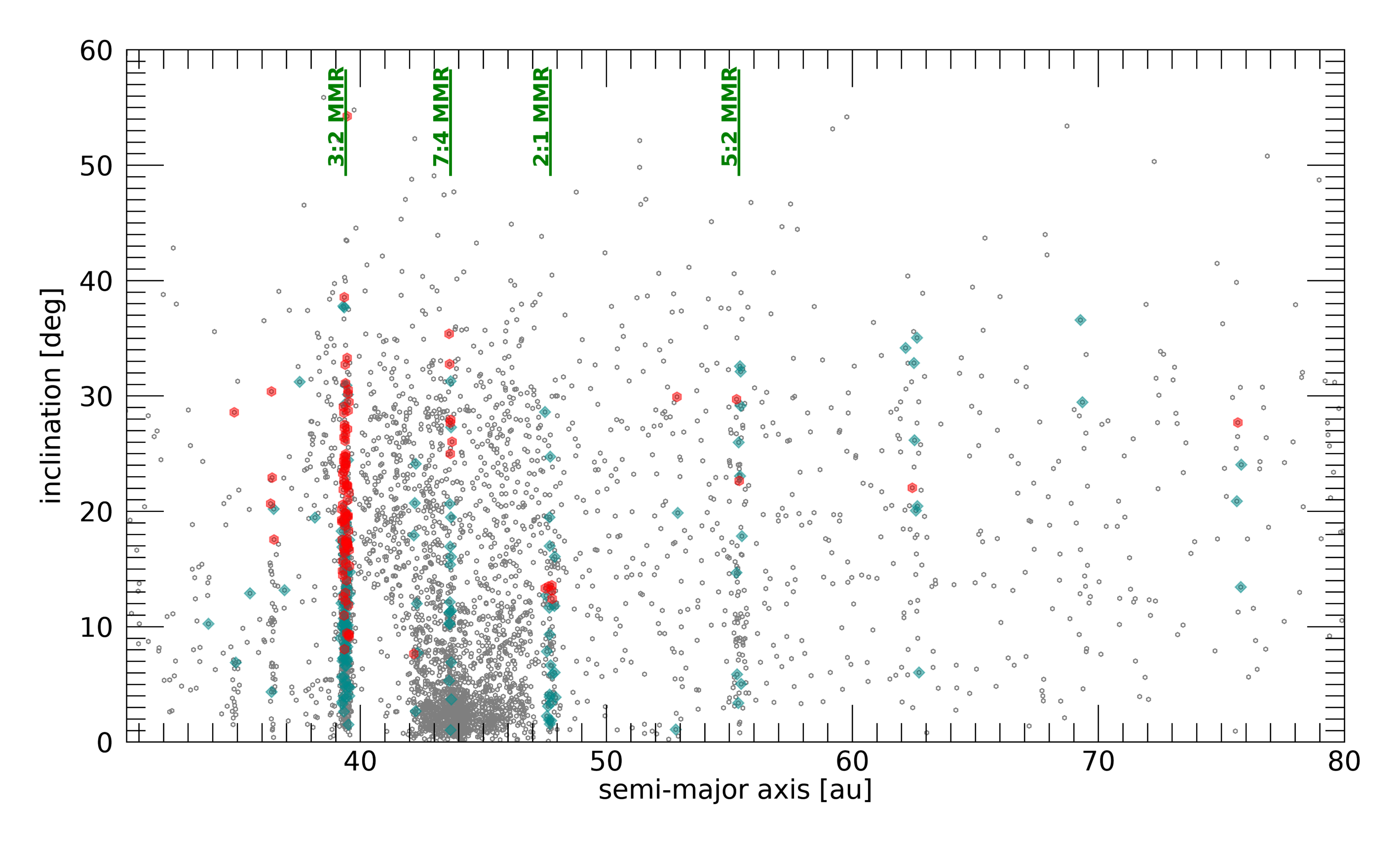}
\caption{\label{fig:tno-all-and-inc-mmr-sma-inc}
Distribution of TNOs in the $(a,I)$ plane. Grey dots denote non-resonant TNOs, coloured ones stand for inclination-type resonant TNOs: red dots indicate TNOs that are engaged in the given MMR throughout the whole $10^8$-year-long time span of the numerical integration, cyan dots mark TNOs that are captured temporarily in the given MMR. Vertical green lines highlight the location of MMRs with more than 10 members.
}
\end{figure}

As we have already mentioned in the Introduction, the trans-Neptunian region includes a remarkably extended range of semi-major axes. In our investigations, we consider objects orbiting between 30.1 AU and 2000 AU. Figures~\ref{fig:tno-all-and-ecc-mmr-sma-ecc}-\ref{fig:tno-all-and-inc-mmr-sma-inc} display the (number) density of these TNOs in the $(a,e)$ and $(a,I)$ planes. The region $a>80$ AU is excluded from the figures, because of the very limited number of known bodies. Furthermore, we do not distinguish between the classical dynamical classes, only between the resonant and non-resonant TNOs (see the different colours).

\begin{figure}
\centering
\includegraphics[width=0.45\linewidth]{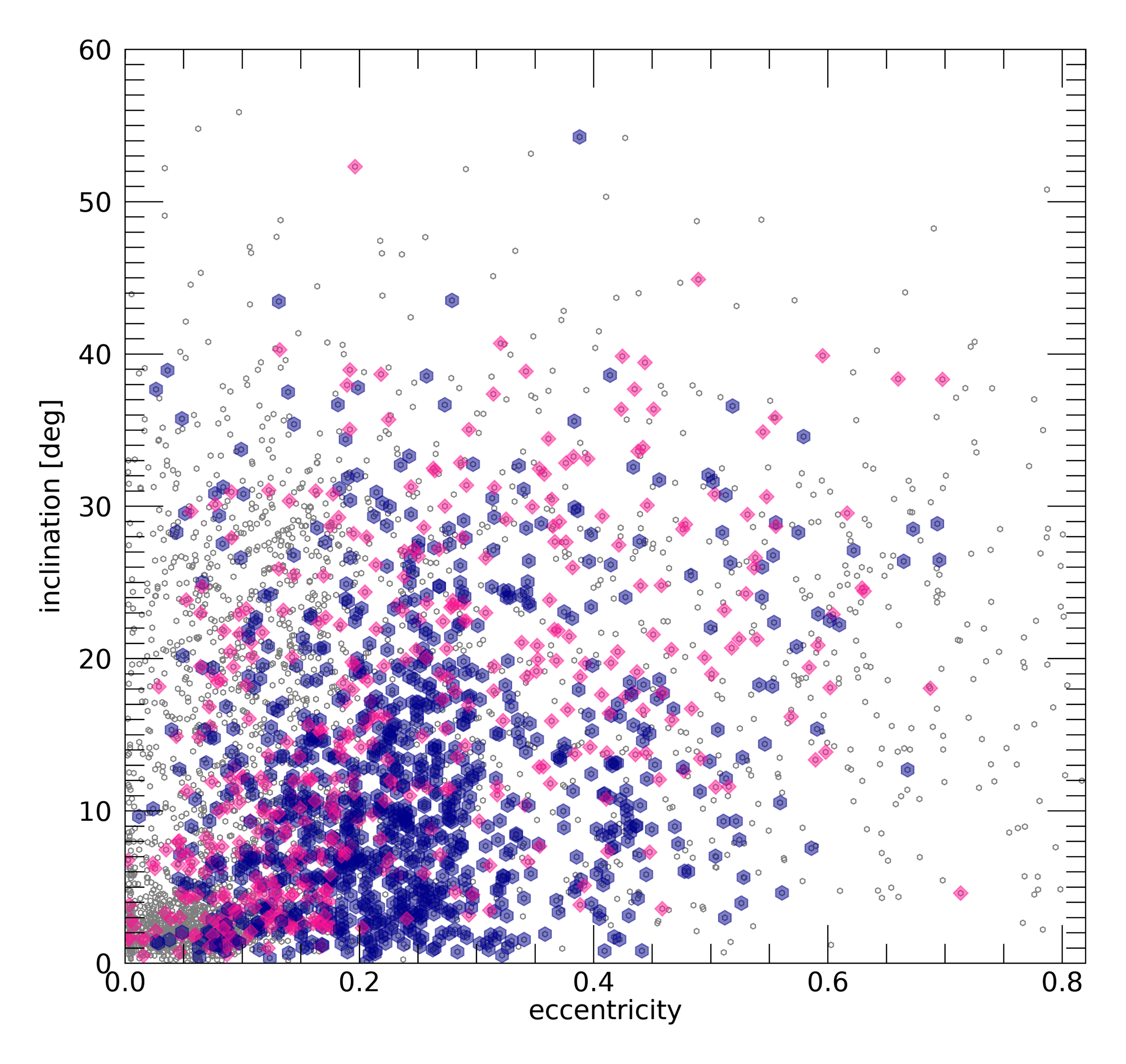}
\includegraphics[width=0.45\linewidth]{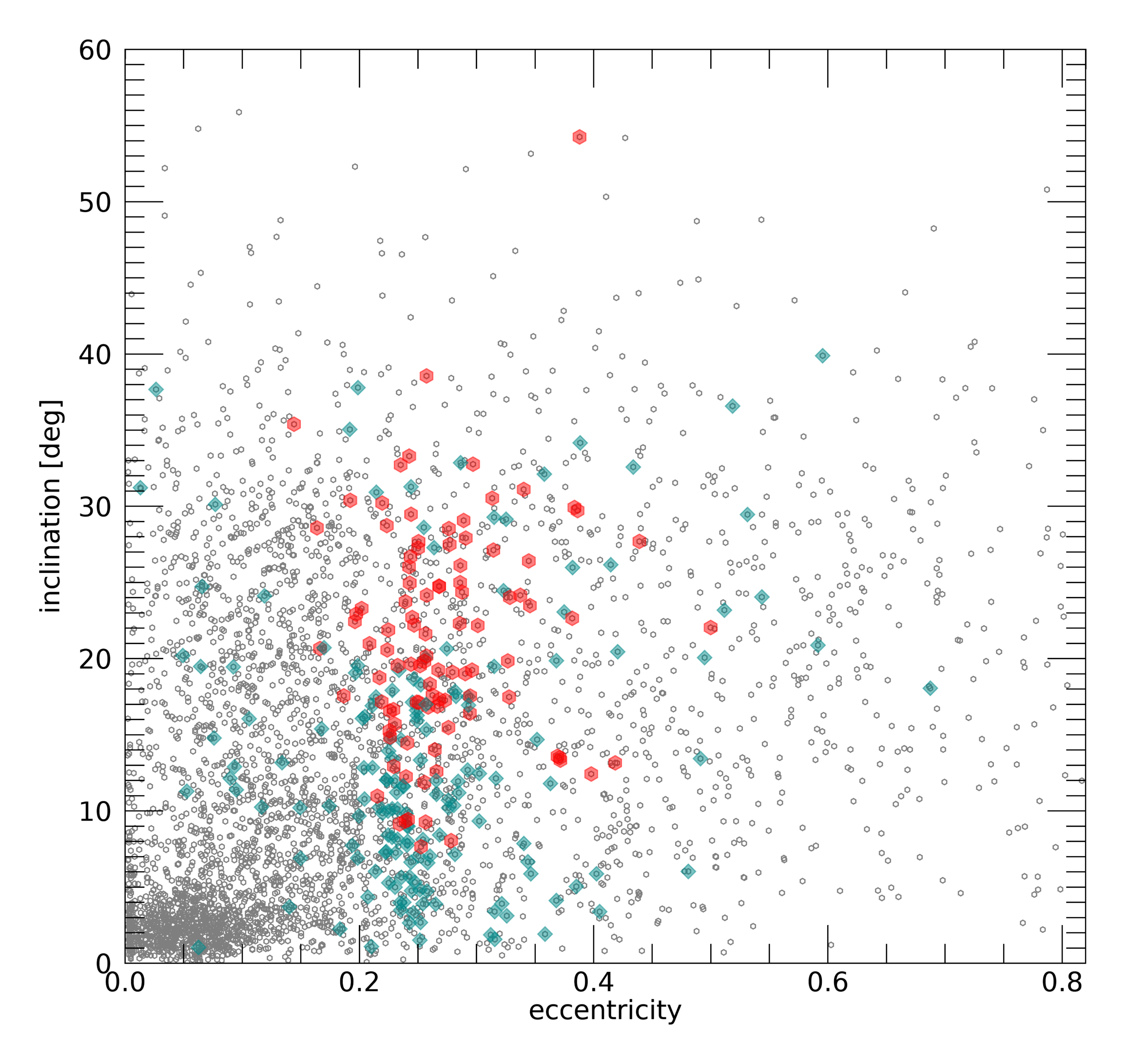}
\caption{\label{fig:tno-all-and-ecc-inc-mmr-ecc-inc}
Distribution of TNOs in the $(e,I)$ plane. Left panel: TNOs involved in eccentricity-type MMRs - the colour coding is the same as in Figure~\ref{fig:tno-all-and-ecc-mmr-sma-ecc}. Right panel: TNOs involved in inclination-type MMRs - the colour coding is the same as in Figure~\ref{fig:tno-all-and-inc-mmr-sma-inc}.
}
\end{figure}

Figure~\ref{fig:tno-all-and-ecc-mmr-sma-ecc} shows in the $ (a,e) $ plane, apart from the orbital distribution of all known TNOs, their resonant/non-resonant character, too. Particularly, the non-grey dots indicate TNOs involved in pure eccentricity-type outer MMRs with Neptune. The blue dots mark TNOs whose critical argument librates throughout the $10^8$-year-long numerical integration (this type of resonance will henceforth be referred to as a long-term MMR). In addition, indicated by pink dots, there is also a significant population of TNOs that are only temporarily locked in a given MMR, i.e., whose critical argument exhibits only a short-term period of libration. We also display by green vertical lines the resonant semi-major axes of MMRs with more than 10 members. We note here that, in this case, the mean-motion commensurability (MMC) would be a more precise designation, referring to those grey-coloured TNOs who are non-resonant even though the ratio of their mean motions are close to the ratio of the given MMR. That is, their critical argument does not librate. It is possible that these objects are in a "post-MMR" phase of their evolution, after a rejection from the exact MMR.

Figure~\ref{fig:tno-all-and-inc-mmr-sma-inc} displays the distribution of TNOs in the $(a, I)$ plane. The inclination-type resonant MMRs are depicted by red and cyan colours. The former ones mark TNOs of long-term MMRs, whereas the latter indicates short-term captures in resonances. Again, inclination-type MMRs of more than 10 members are highlighted by green vertical lines. Comparing the two figures, it is apparent that the 3:2 MMR is the most populated one, both in terms of eccentricity- and inclination-type resonances.

In addition to the "traditional" distribution plots, the initial orbital elements of the TNOs are indicated in the $(e, I)$ plane, too (see Figure~\ref{fig:tno-all-and-ecc-inc-mmr-ecc-inc}). The eccentricity- and inclination-type resonances are shown separately in the left and right panels, respectively. The following observations can be made with regard to the distributions. The most striking finding is that TNOs engaged in inclination-type resonances (right panel) are characterized by eccentricities $ e \sim 0.2-0.3 $. It is also discernible that the short- and long-term types are well-separated in the inclination, i.e., the small bodies locked in long-term inclination-type MMRs are positioned above 10 degrees of inclination. In the case of the TNOs in eccentricity-type resonances (left panel), one observes that in the eccentricity range $ 0.2 < e < 0.3 $, for lower values of inclination, almost all resonant objects exhibit a stable libration over the entire 100 million-year-long time span.

We emphasize that the above figures show two different subtypes of mean-motion resonances in contrast to previous works where only eccentricity-type MMRs were considered. Although certain studies mention minor bodies involved in inclination-type resonances, such comprehensive surveys have not been carried out so far. In the following subsection, we review in detail the characteristics of the two subtypes by providing examples from our TNO sample.

\subsection{Classification of TNOs involved in MMRs}

\begin{figure}
\centering
\includegraphics[width=0.95\linewidth]{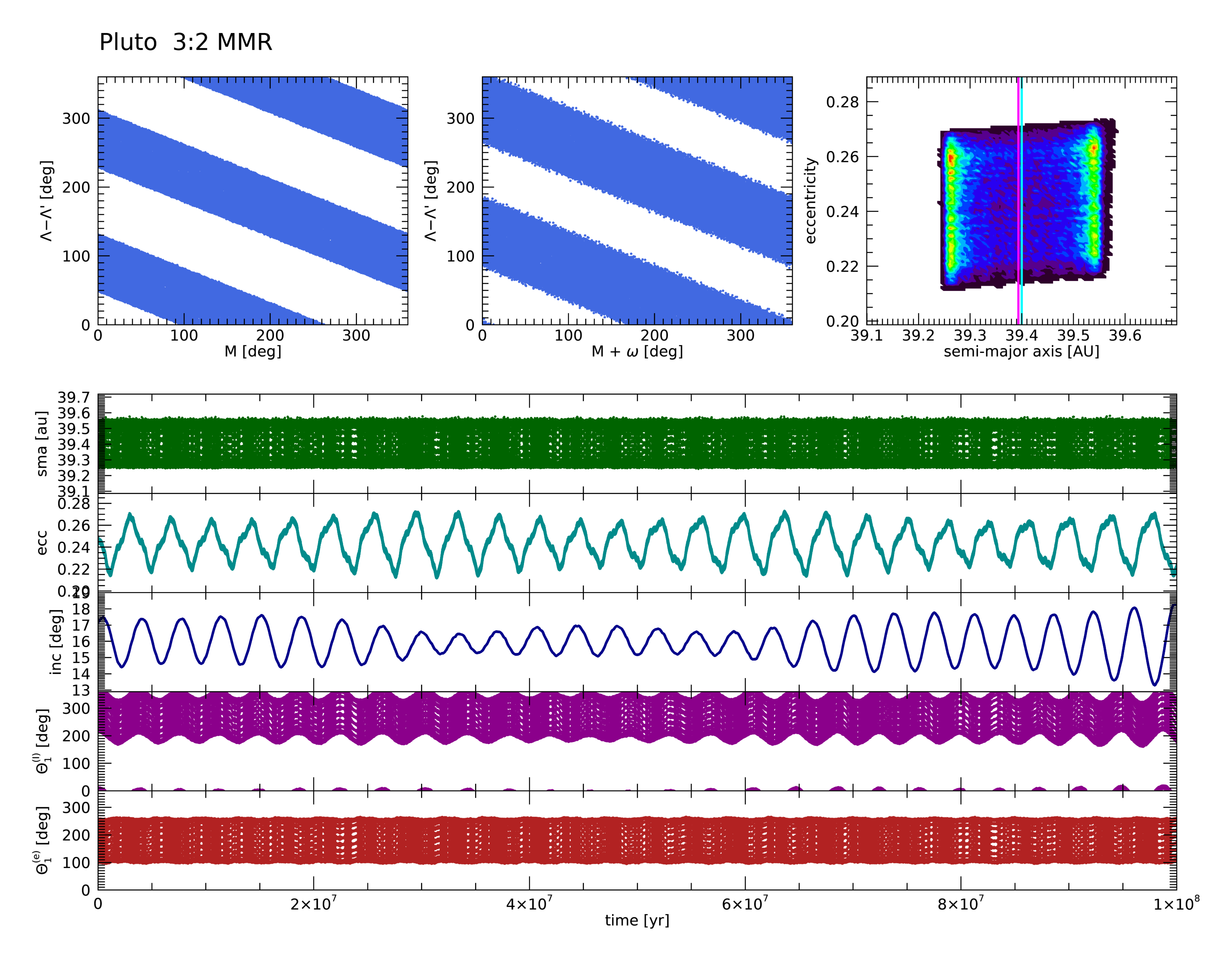}
\caption{\label{fig:3-2-MMR-Both-Pluto}
The dynamics of the dwarf planet Pluto engaged in a $3:2$ MMR with Neptune. The first two panels in the uppermost row show the $\lambda-\lambda^\prime$ vs $M$ and $\lambda-\lambda^\prime$ vs $M+\omega$ planes used to determine the type, order, and degree of the resonance by applying the FAIR method. The third panel reveals the region covered by Pluto in the $(a,e)$ plane during the whole length of the numerical integration. The lower panels display, respectively, the time evolution of the orbital elements $a$, $e$, $I$, and that of the critical arguments of the inclination- and eccentricity-type MMRs $\Theta_1^{(I)}$ and $\Theta_1^{(e)}$.
}
\end{figure}

\begin{figure}
\centering
\includegraphics[width=0.95\linewidth]{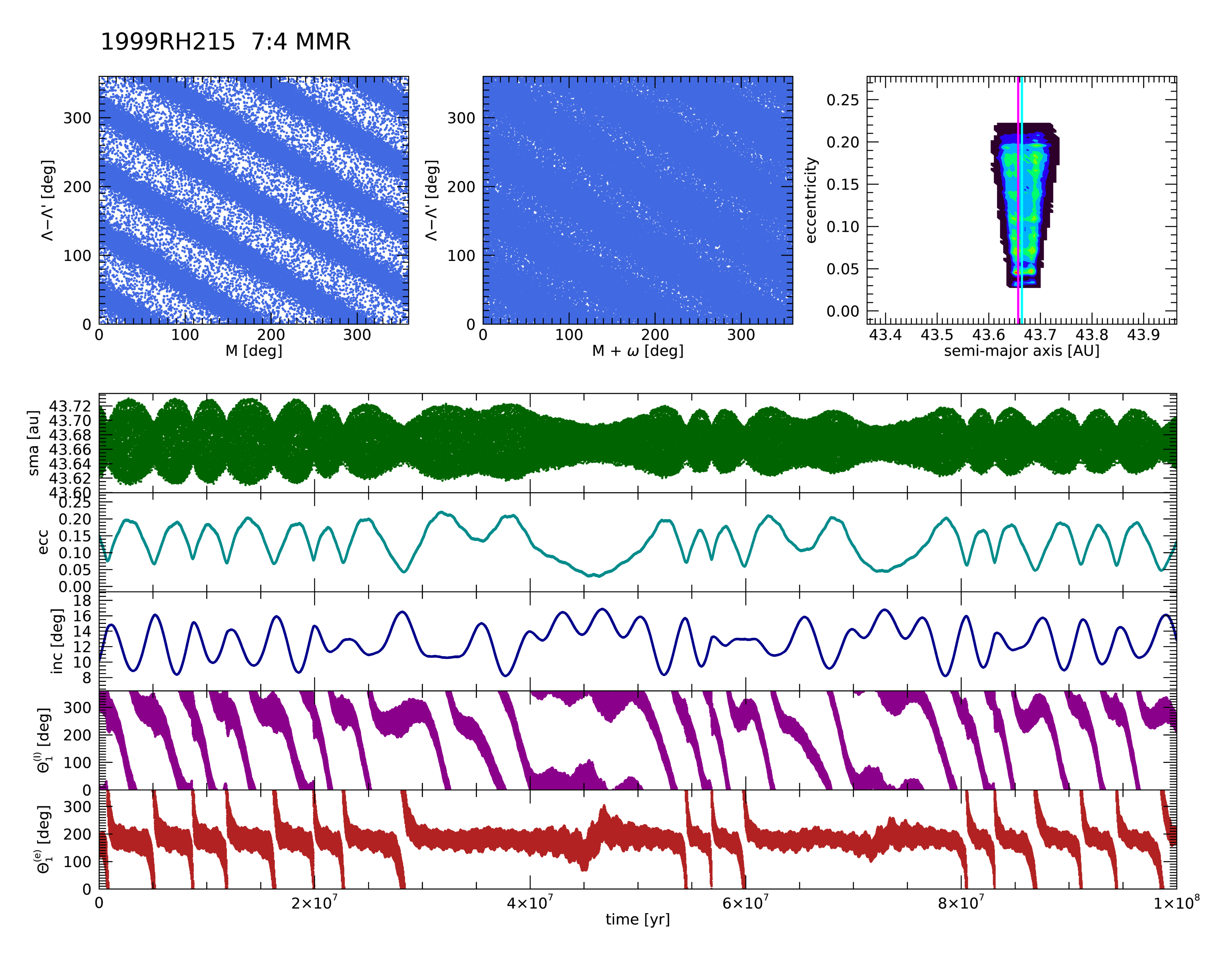}
\caption{\label{fig:7-4-MMR-Both-1999RH215}
The dynamics of the asteroid 1999\,RH$_{215}$ engaged in a $7:4$ MMR with Neptune. The structure of the figure is the same as of Figure~\ref{fig:3-2-MMR-Both-Pluto}. Note the episodical libration of the critical arguments $\Theta_1^{(I)}$ and $\Theta_1^{(e)}$.
}
\end{figure}

As we mentioned earlier, the FAIR method allows determining whether an object is involved in an MMR with a given planet, without any a priori assumption on its dynamics. Our aim was to identify the various resonances among the $\sim 4200$ TNOs of our sample. To do so, we numerically integrated the equations of motion for each individual TNO for a time span of $10^8$ years in a barycentric coordinate system and calculated the barycentric orbital elements, which then enabled the utilisation of the FAIR method. To illustrate how it works, we present our results in the cases of some intriguing TNOs.

A classic example can be seen in Figure~\ref{fig:3-2-MMR-Both-Pluto}. It shows some dynamical properties of the dwarf planet Pluto. The three panels in the first row display plots of $\lambda - \lambda^{\prime}$ vs $M$, $\lambda - \lambda^{\prime}$ vs $M+\omega$, and $a$ vs $e$, respectively. In the first two panels, used to apply the FAIR method, three stripes appear, with $ p=1 $ and $ q=2 $ intersections with the horizontal and vertical axes, respectively. This clearly indicates that Pluto is locked both in an eccentricity- and in an inclination-type MMR with Neptune, of the ratio $3:2$. The third panel here illustrates the excursion of the semi-major axis and eccentricity values of Pluto in the $(a,e)$ plane. The lighter the colouring of the panel, the more frequent the given region of the phase space section. The four lower panels show the evolution of $a$, $e$, $I$, $\Theta_1^{(I)}$, and $\Theta_1^{(e)}$ as functions of time. The latter two, i.e., the critical arguments of the inclination- and eccentricity-type MMRs, respectively, are obtained by adopting the resonant parameters $p=1$ and $q=2$ (see Equations~\eqref{eq:crit_arg_e} and \eqref{eq:crit_arg_inc}).

Besides differentiating between the eccentricity- and inclination-type MMRs, we should also distinguish between the cases where the corresponding critical argument librates during the whole integration time, and where its libration is only temporary. In the latter case, the libration can happen multiple times, that is, intervals of libration and circulation can alternate several times during the numerical integration. This behaviour is shown in Figure~\ref{fig:7-4-MMR-Both-1999RH215}, where the dynamical behaviour of the asteroid 1999\,RH$_{215}$, of a $7:4$ MMR with Neptune, is displayed. One sees in the first two panels of the first row that in between the stripes there are scattered points, too. This, by applying the FAIR method, implies the temporary character of the critical arguments' libration. Both $\Theta_1^{(I)}$ and $\Theta_1^{(e)}$ feature shorter or longer periods of libration as well as of circulation. It is noteworthy that the libration of the two critical arguments is not synchronized; it can happen, for instance, that during the libration of $\Theta_1^{(e)}$, the critical argument $\Theta_1^{(I)}$ circulates.

The two examples discussed above, demonstrate the use of the FAIR method. In a similar - though semi-automatized - manner, all small bodies of our study were examined in terms of MMRs. Our results are summarized in Tables~\ref{tab:numofetypemmr}-\ref{tab:numofitypemmr}. Out of the entire sample of 4121 TNOs, 1359 small bodies are found to be engaged - either in a short- or in a long-term - eccentricity-type resonance (Table~\ref{tab:numofetypemmr}), while the number of those involved in an inclination-type MMR is 260 (Table~\ref{tab:numofitypemmr}). In the tables, only resonances with more than 10 members are shown, examples of the "more exotic" MMRs are given below. It is also important to note that out of the 260 TNOs exhibiting an inclination-type resonance, 257 feature an eccentricity-type MMR, too (see e.g. Figures~\ref{fig:3-2-MMR-Both-Pluto} and \ref{fig:7-4-MMR-Both-1999RH215}). We generalize this observation and claim that those TNOs found to be in inclination-type resonances are most likely to show an eccentric-type resonance as well. This property will be discussed later in Section~\ref{sec:summa}.

\begin{deluxetable}{ccccc}
\tablecaption{Distribution of the eccentricity-type MMRs, hosting more than 10 TNOs.\label{tab:numofetypemmr}}
\tablewidth{0pt}
\tablehead{
\colhead{\shortstack{e-type\\MMR}} & \colhead{SMA} & \colhead{\shortstack{No. of TNOs\\of the MMR}} & \colhead{\shortstack{No. of\\LT TNOs}}  & \colhead{\shortstack{No. of\\ST TNOs}} }
\startdata
 5:4  &   34.890  &  16   &  14 (100\%)   &    2 (50\%)  \\       
 4:3  &   36.424  &  45   &  40 (77\%)   &    5 (60\%)  \\        
 3:2  &   39.400  & 505   & 477 (79\%)   &   28 (46\%)  \\        
 5:3  &   42.267  &  94   &  53 (86\%)   &   41 (56\%)  \\        
 7:4  &   43.664  & 145   &  53 (77\%)   &   92 (69\%)  \\        
 9:5  &   44.492  &  19   &   5 (60\%)   &   14 (35\%)  \\        
11:6  &   45.040  &  16   &   2 (100\%)   &   14 (42\%)  \\        
 2:1  &   47.730  & 128   &  89 (76\%)   &   39 (61\%)  \\        
 7:3  &   52.895  &  21   &  16 (62\%)   &    5 (100\%)  \\        
 5:2  &   55.385  &  88   &  63 (82\%)   &   25 (72\%)  \\        
 8:3  &   57.820  &  12   &   4 (75\%)   &    8 (87\%)  \\        
 3:1  &   62.543  &  30   &  19 (73\%)   &   11 (90\%)  \\        
 7:2  &   69.313  &  19   &  13 (84\%)   &    6 (66\%)  \\        
 4:1  &   75.766  &  11   &   5 (80\%)   &    6 (83\%)
\enddata
\tablecomments{$1^{\mathrm{st}}$ column: ratio of the eccentricity-type MMR; $2^{\mathrm{nd}}$ column: semi-major axis at the centre of the resonance; $3^{\mathrm{rd}}$ column: number of TNOs in the MMR of the first column; $4^{\mathrm{th}}$ column: number of TNOs in the MMR of the first column with long-term libration; $5^{\mathrm{th}}$ column: number of TNOs in the MMR of the first column with short-term libration. In the last two columns, the values in the parentheses give the percentage of those TNOs whose OCC number is less than 5, i.e., the greater the number, the more reliable the data is.}

\end{deluxetable}

\begin{deluxetable}{ccccc}
\tablecaption{Distribution of the inclination-type MMRs, hosting more than 10 TNOs.\label{tab:numofitypemmr}}
\tablewidth{0pt}
\tablehead{
\colhead{\shortstack{I-type\\MMR}} & \colhead{SMA} & \colhead{\shortstack{No. of TNOs\\of the MMR}} & \colhead{\shortstack{No. of\\LT TNOs}}  & \colhead{\shortstack{No. of\\ST TNOs}} }
\startdata
 3:2 &   39.400 & 160 & 77 (90\%) & 83 (84\%) \\         
 7:4 &   43.664 & 23  & 6  (83\%) & 17 (64\%) \\         
 2:1 &   47.730 & 27  & 5  (80\%) & 22 (90\%) \\         
 5:2 &   55.385 & 12  & 2  (100\%) & 10 (90\%)        
\enddata
\tablecomments{See Table~\ref{tab:numofetypemmr} for the explanation of the columns.}
\end{deluxetable}

By observing Table~\ref{tab:numofetypemmr}, one finds that among the detected TNOs the most populated resonance is the eccentricity-type 3:2 MMR, which covers nearly 40\% of the entire resonant population. (In this population, 391 objects (77\%) have an OCC less than 5, thus even if we were to omit the complement 23\%, the 3:2 MMR would still remain the most populous one.) In addition to the total number of resonant TNOs (third column), we provide the number of those involved in long- (fourth column) and short-term MMRs (fifth column), respectively, too. Again, in this respect, the 3:2 MMR is the one containing most of the long-term types; almost 95\% of the population. It is an interesting finding that the second most populated resonance is neither a first- nor a second-order but a third-order one, namely the 7:4 MMR (even if we take into account that 28\% of the resonance is characterized by an OCC $> 5 $). However, this resonance is mostly dominated by TNOs with critical arguments that do not librate throughout the entire time span. The third-place winner MMR is, again, a first-order type, more precisely, the 2:1 MMR. In this first-order resonance, more than 70\% of the bodies librate stably throughout the whole integration time, in contrast to the previously mentioned third-order 7:4 resonance, where this ratio was just the opposite. Yet evidently, the distribution of the TNOs of long- and short-term resonances is influenced by several factors: the 5:2 MMR is a third-order type once again, but with a percentage of more than 70\% of stable, long-term MMRs. 
Nontheless, in general, it can be said that the first-order resonances are "stronger" than their higher-order counterparts. This fact is supported by Table~\ref{tab:numofetypemmr}: the portion of the long-term MMRs in the case of the first-order resonances is mainly over 2/3. In contrast, higher-order resonances are less numerous, on the one hand, and the proportion of TNOs in short-term MMRs increases significantly, on the other. The aforementioned 5:2 as well as e.g. the fifth-order 7:2 MMRs can be regarded as exceptions, with remarkably more bodies in long-term resonances. 
Future measurements will clarify whether the above findings concerning the resonant distribution of the TNOs can be solely explained by the dynamics of the region, for the moderate number of inhabitants in the more distant resonances might be the consequence of observational biasses, too. In this study, we try to overcome this puzzle by presenting dynamical maps of the $ 30-80 $ AU region of the trans-Neptunian space (see Section \ref{sec:dyn-map}).

\begin{figure}
\centering
\includegraphics[width=0.95\linewidth]{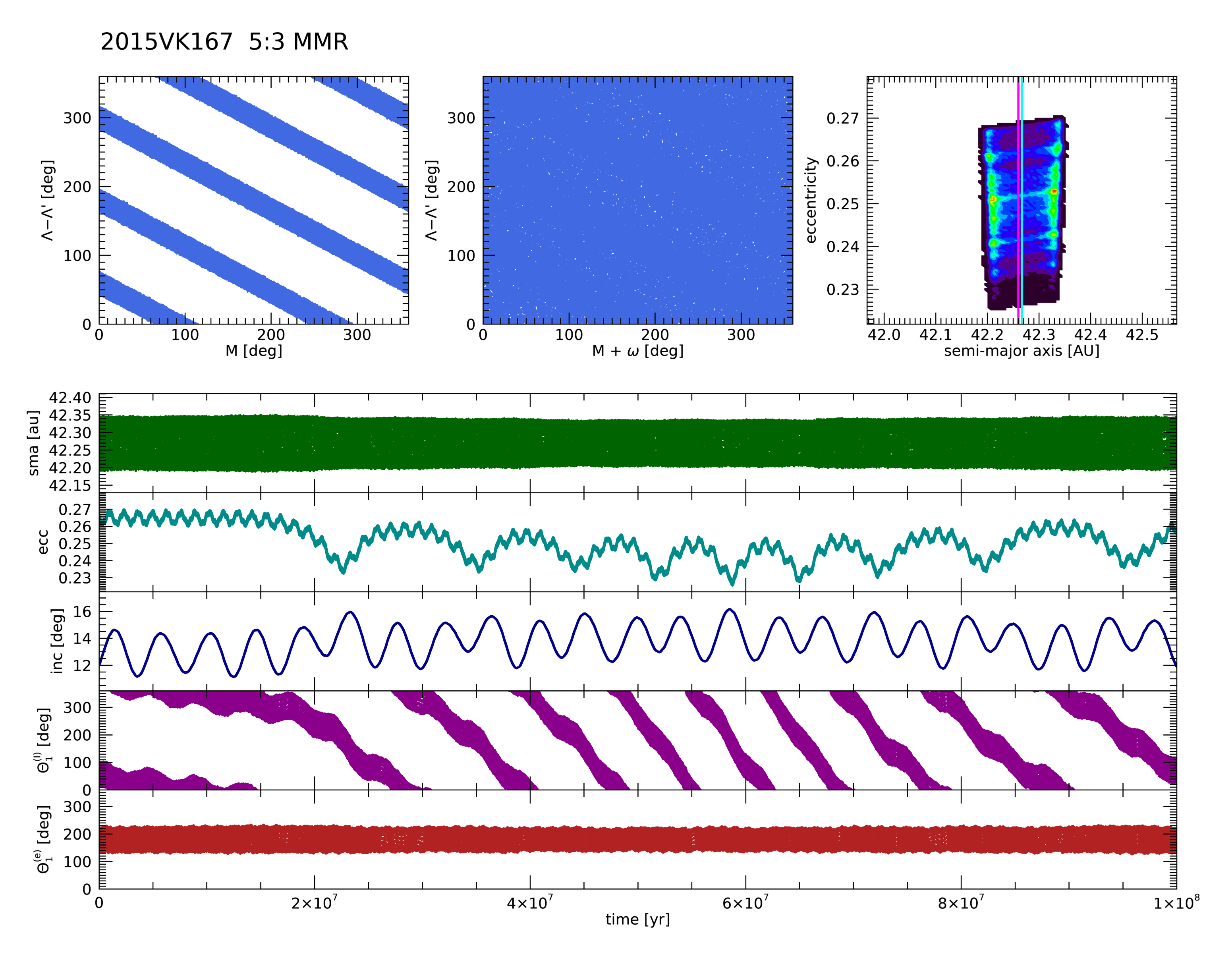}
\caption{\label{fig:5-3-MMR-2015VK167}
The dynamics of the asteroid 2015\,VK$_{167}$ engaged in a $5:3$ MMR with Neptune. The structure of the figure is the same as of Figure~\ref{fig:3-2-MMR-Both-Pluto}. In this case, $\Theta_1^{(I)}$ librates only temporarily, while $\Theta_1^{(e)}$ librates during the whole numerical integration.
}
\end{figure}

Let us now turn to the discussion of Table~\ref{tab:numofitypemmr}, summarizing the abundance of the inclination-type MMRs. (We emphasise again that among the 260 TNOs trapped in inclination-type MMRs, 257 objects are also involved in eccentricity-type MMRs.) The first observation of the table is that among the inclination-type MMRs, the proportion of the short-term types is remarkably larger than in the case of the eccentricity-type resonances. It is also noteworthy that all the TNOs of long-term inclination-type MMRs manifest a long-term eccentricity-type MMR, too. That is, the long-term libration of $\Theta_1^{(I)}$ is not possible if $\Theta_1^{(e)}$ librates only temporarily, only the other way around. Figure~\ref{fig:5-3-MMR-2015VK167} provides an example for the case of a long-term librating $\Theta_1^{(e)}$ and an episodically circulating $\Theta_1^{(I)}$.

\subsection{TNOs with peculiar dynamical properties}

\begin{figure}
\centering
\includegraphics[width=0.9\linewidth]{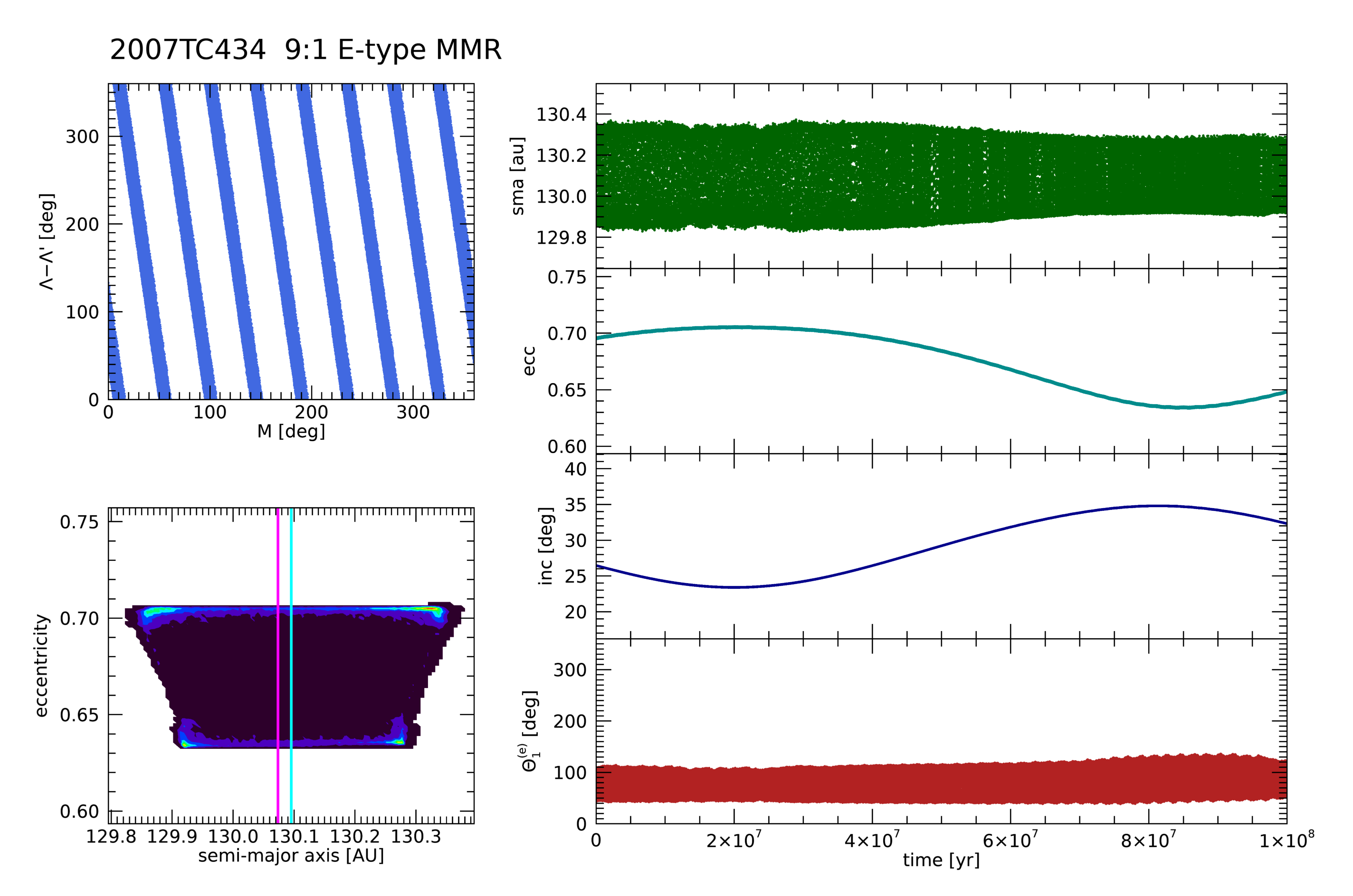}
\caption{\label{fig:9-1-MMR-2007TC434}
The dynamics of the asteroid 2007\,TC$_{434}$ engaged in a $9:1$ MMR with Neptune. The structure of the shown panels is the same as in Figure~\ref{fig:3-2-MMR-Both-Pluto}; however, in this case, only $\Theta_1^{(e)}$ librates; therefore, neither the panel to identify the inclination-type MMR nor the time evolution of $\Theta_1^{(I)}$ are displayed. This object is involved only in an eccentricity-type MMR.
}
\end{figure}

In what follows, we present some examples of those trans-Neptunian objects that we consider as peculiar in terms of resonant behaviour. 
The purpose of the present subsection is to illustrate the dynamical diversity of the TNOs. However, we must draw the reader's attention once again to the fact that since the initial coordinates of the objects stem from the best fit of the observations taken place so far, with the future refinement of the input data, the whole evolution of an individual TNO might alter. Consequently, the extremes (largest, furthest, faintest, etc.) of the TNO population could also be affected.

As a first example, we show in Figure~\ref{fig:9-1-MMR-2007TC434} the most distant resonant TNO, the asteroid 2007\,TC$_{434}$, which is involved in a $9:1$ eccentricity-type MMR with a semi-major axis of 130.1 AU. Despite its great distance and the high order of its MMR, the asteroid exhibits surprisingly regular dynamics.

\begin{figure}
\centering
\includegraphics[width=0.9\linewidth]{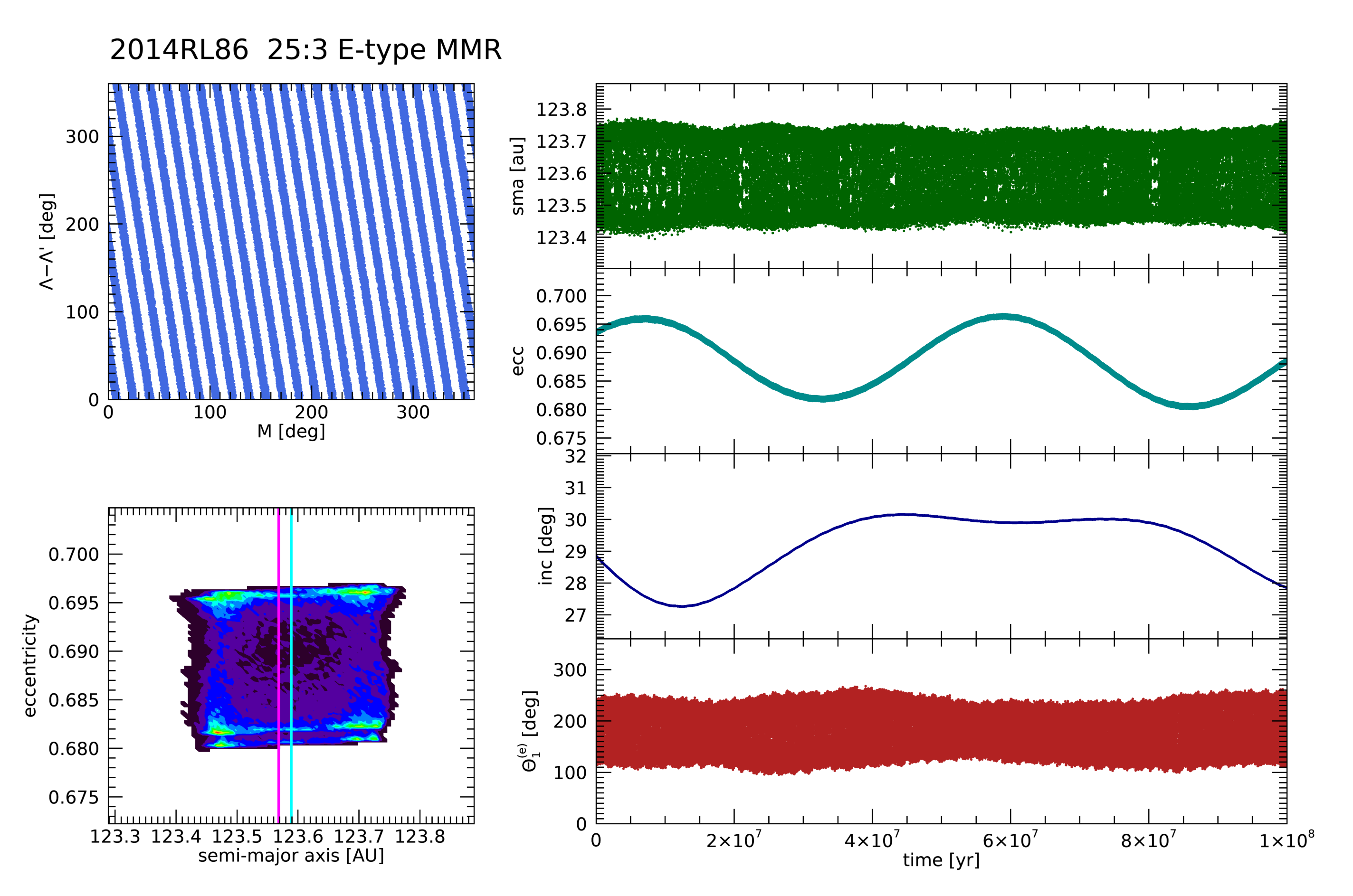}
\caption{\label{fig:25-3-MMR-2014RL86}
The dynamics of the asteroid 2014\,RL$_{86}$ engaged in a $25:3$ eccentricity-type MMR with Neptune. The structure of the figure is the same as of Figure~\ref{fig:9-1-MMR-2007TC434}.
}
\end{figure}

The next example, presented in Figure~\ref{fig:25-3-MMR-2014RL86}, is the highest-order resonance of our sample. The TNO 2014\,RL$_{86}$ orbits the Sun at $ a=123.6$ AU and is stably locked in the 22nd-order eccentricity-type $25:3$ MMR. Besides this object, there are several other TNOs involved in very high-order MMRs with critical arguments yielding libration for the whole integration time span.

\begin{figure}
\centering
\includegraphics[width=0.9\linewidth]{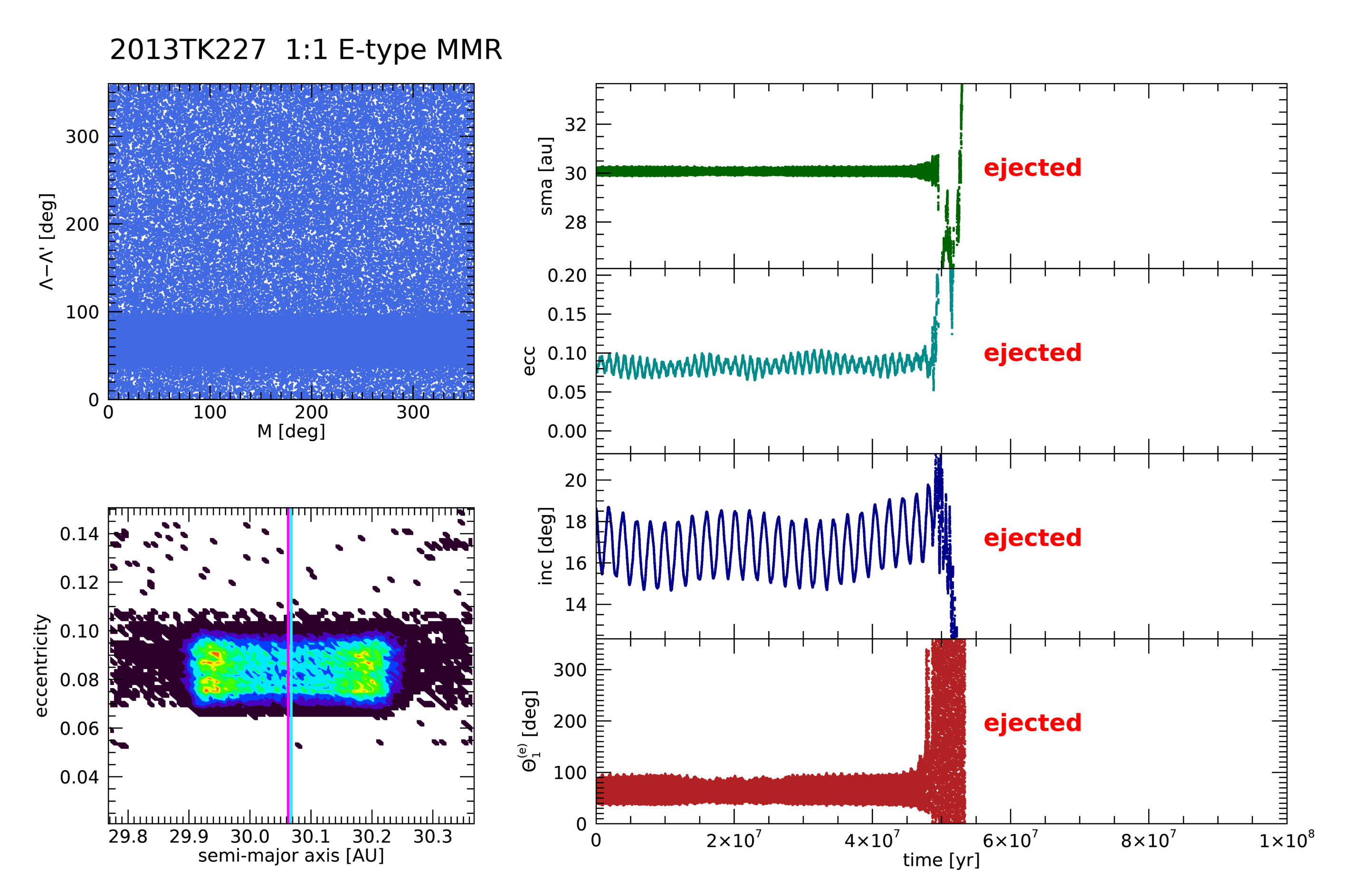}
\caption{\label{fig:1-1-MMR-2013TK227}
The dynamics of the asteroid 2013\,TK$_{227}$, a temporary Trojan of Neptune. The structure of the figure is the same as of Figure~\ref{fig:9-1-MMR-2007TC434}.
}
\end{figure}

\begin{figure}
\centering
\includegraphics[width=0.9\linewidth]{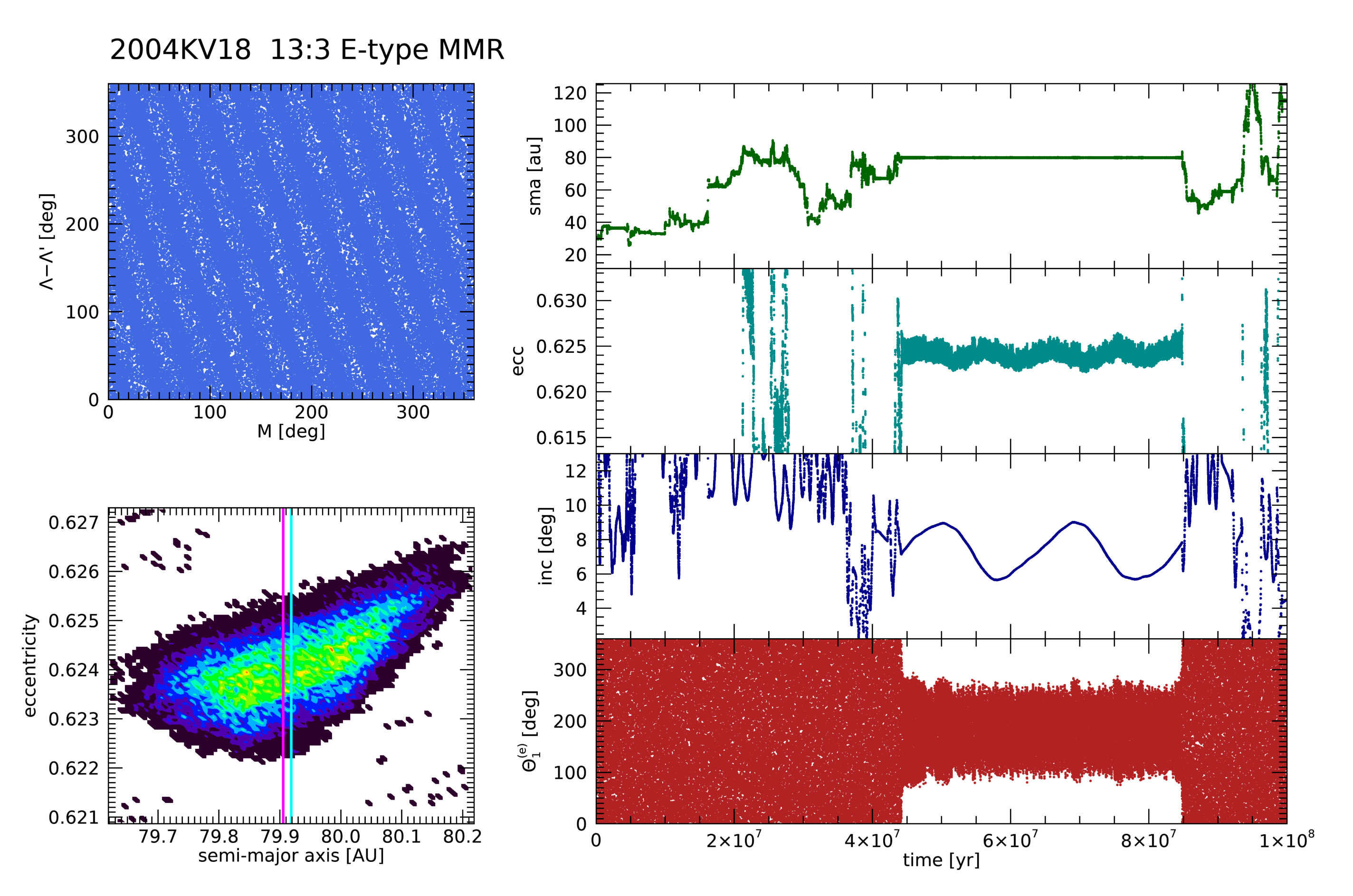}
\caption{\label{fig:13-3-MMR-2004KV18}
The dynamics of the asteroid 2004\,KV$_{18}$, identified previously as a temporal Trojan of Neptune. Our long-term numerical integration shows that this object is captured temporarily in the $13:3$ eccentricity-type MMR spending there about $4\times 10^7$ years. The structure of the figure is the same as of Figure~\ref{fig:9-1-MMR-2007TC434}.
}
\end{figure}

After the presentation of the record holder TNOs, let us move on to the discussion of an important asteroid family, namely, the Trojan asteroids of Neptune. First, in Figure~\ref{fig:1-1-MMR-2013TK227}, we present the dynamical behaviour of the TNO 2013\,TK$_{227}$. Its discovery in 2013 was followed by its classification as a Neptune-Trojan in 2021 after the refinement of its orbital elements. It occupies the L$_4$ Lagrange point of the Sun-Neptune system, with an OCC value of 4, that is, the longitude runoff per decade is less then 6.4 arc minutes.
(We note here that by using the FAIR method, Trojan-like objects (involved in the $1:1$ MMR) can be identified even if they possess drastically changing orbital elements.) In the FAIR panel of Figure~\ref{fig:1-1-MMR-2013TK227}, the temporary capture of the object into the $1:1$ MMR is visible (see the horizontal stripe around $60^{\circ}$ in the $( \lambda - \lambda^{\prime}, M)$ plane). The temporal character of this resonance is indicated by the set of scattered points around the horizontal stripe.

The next Trojan asteroid of Neptune that we consider is the 2004\,KV$_{18}$ (Figure~\ref{fig:13-3-MMR-2004KV18}). This object was identified previously as "Neptune's temporary Trojan" located at the L$_5$ Lagrange point, again, with an OCC value of 4. The "Trojan period" of the TNO is not discernible in the figure since it lasts for only $\sim 10^4$ years at the beginning of the integration. However, after having escaped from the $1:1$ MMR, it is captured in the distant $13:3$ MMR at 79.92 AU, followed by a relatively long period of chaotic wandering. 
According to our simulations, it spends nearly $4\times 10^7$ years in the latter MMR before being rejected from that, too. 

The above-mentioned Neptune Trojans show peculiar motion that results either in a temporal capture in different resonances or in an escape from the system. These substantially different fates of the asteroids stem from the highly nonlinear nature of the TNO dynamics. A common technique to investigate irregular motion in dynamical astronomy is to study a large number of clones scattered in the phase in the vicinity of the object in question. The choice of the initial conditions of the ensemble of virtual particles is somewhat arbitrary. The details of our choice for the numerical setup of the test particles will be given in Sec.~\ref{sec:dyn-map}. The question might then arise whether the fictitious clones provide suitable substitutes for the observational data of the real TNO obtained at different epochs. The answer to this question, due to its theoretical nature, is postponed to a future project.

\begin{figure}
\centering
\includegraphics[width=0.9\linewidth]{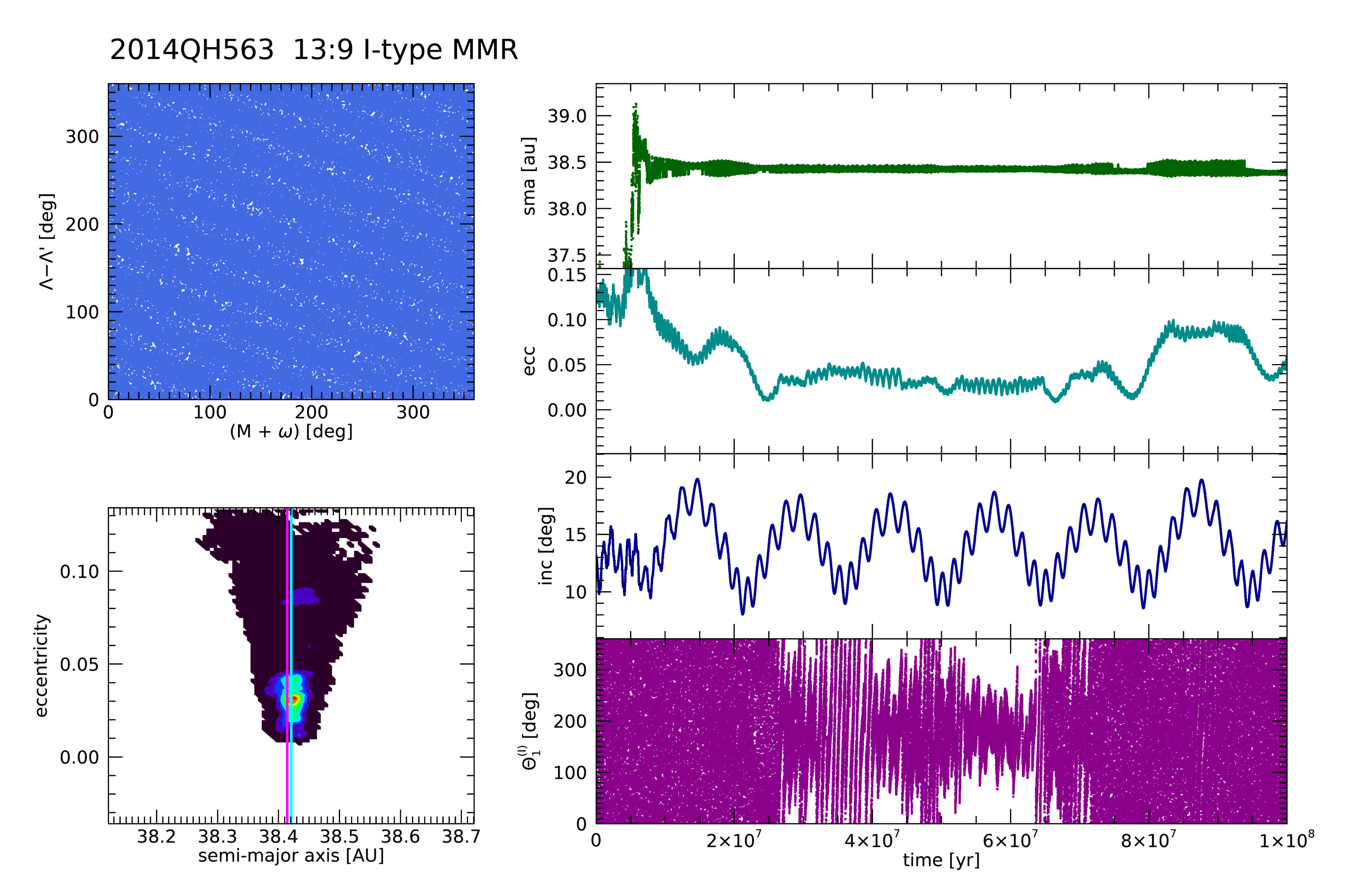}
\caption{\label{fig:13-9-MMR-2014QH563}
The dynamics of the asteroid 2014\,QH$_{563}$ engaged in a $ 13:9 $ inclination-type MMR with Neptune. The structure of the figure is the same as of Figure~\ref{fig:9-1-MMR-2007TC434}, but showing this time the FAIR panel and the time evolution of the critical argument corresponding to an inclination-type MMR.
}
\end{figure}

Our last example of the by some means interesting TNOs is the case of the asteroid 2014\,QH$_{563}$ that yields a pure inclination-type $13:9$ MMR. Its uniqueness lies in the fact that it is one of those 3 TNOs that show only an inclination-type MMR but not an eccentricity-type one. The critical argument $\Theta_1^{(I)}$ manifests episodic libration, while $\Theta_1^{(e)}$ circulates during the whole time of the numerical integration. The dynamical behaviour of this object is shown in Figure~\ref{fig:13-9-MMR-2014QH563}; however, the time evolution of $\Theta_1^{(e)}$ is not displayed, only that of $\Theta_1^{(I)}$. Besides the temporary nature of the libration of the critical argument $\Theta_1^{(I)}$, its further peculiarities are the very irregular variations both in the libration amplitude and in its mean value. Interestingly, during the libration period of $\Theta_1^{(I)}$, the orbital element that suffers the biggest variations is the eccentricity as its value becomes nearly zero. As for the inclination, its variation seems to be quite regular throughout the entire integration time span. In this particular case, it is most probable that those terms become dominant in the perturbing function whose coefficients are related to the inclination of the TNO. This property would support the formation of the inclination-type MMR without the appearance of the eccentricity-type one.

In this subsection, we presented some of the most intriguing TNOs, but there are several other objects that deserve special attention. These bodies will be investigated in a forthcoming paper. In the present work, our primary objective is to give an overview of all known TNOs, whether resonant or not; however, we already hint at our future investigations aiming at a better understanding of the origins of the dynamical evolution of these bodies.

\subsection{Dynamical maps of the trans-Neptunian region}\label{sec:dyn-map}

In the previous subsections, we revealed several interesting phenomena regarding the resonant behaviour of the faraway asteroids in the trans-Neptunian region. For a more detailed understanding of the underlying dynamics, we compiled dynamical maps of test particles sampled between 30 and 80 AU.

The shown quantity on the dynamical maps was first chosen to be the maximum variation of the eccentricities (see Section~\ref{sec:data_and_methods}). When applying this efficient and widely used method, one usually assigns to each $(a,e)$ value (taken from a rectangular grid of the semi-major axis -- eccentricity plane) the corresponding maximum variation of the orbital eccentricity of the object obtained during the entire time span of the numerical integration. We used a similar approach but with certain modifications described in the following. In order to investigate the whole trans-Neptunian region, we placed a large number of test particles on a rectangular grid of the $(a,e)$ parameter plane while the orbital elements of the major planets were kept fixed at their own values of a given epoch. The phase space section, in the case of each dynamical map (see the forthcoming figures), was divided into 100 big grid cells both in the horizontal and vertical directions. This resulted in $10^4$ big grid cells per map. Thereafter, each big grid cell was further divided into a subgrid containing 20 small grid cells, 5 in the horizontal and 4 in the vertical directions. The initial $(a,e)$ values of the fictive particles were taken from this latter and finer subgrid, while the other orbital elements were randomly chosen. To the big grid cell, we assigned the mean value of the maximum variations of the eccentricity of the 20 test particles of the subgrid. In this manner, instead of fixing the angular-like orbital elements to zero (or to some predefined value) - which is the general approach used in previous works -, we introduced some randomness in the simulations, which concept we consider more realistic. This way the dependence of the dynamical map on the orbital positions of the test particles is eliminated, and the dynamical map represents the most dominant dynamical behaviour of the big grid cell. This method is, however, computationally very expensive. To illustrate the number of test particles involved in our simulations, let us consider the case of a dynamical map of $34\leq a \leq 40$ AU and $0\leq e \leq 0.6$. This example yields $2\times 10^5$ test particles altogether, which ensemble was integrated numerically for $2\times 10^5$ years.

\begin{figure}
\centering
\includegraphics[width=0.45\linewidth]{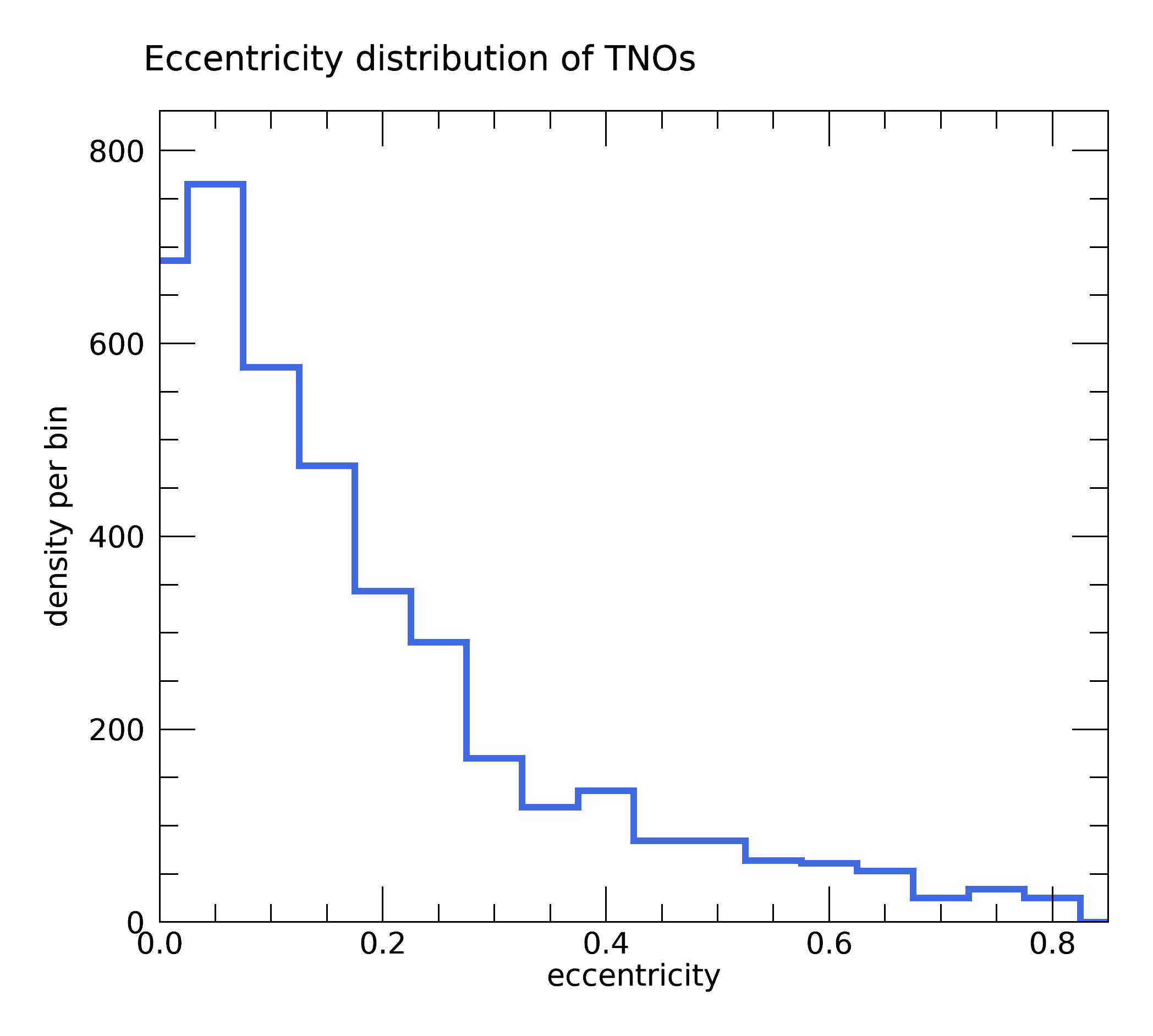}
\includegraphics[width=0.45\linewidth]{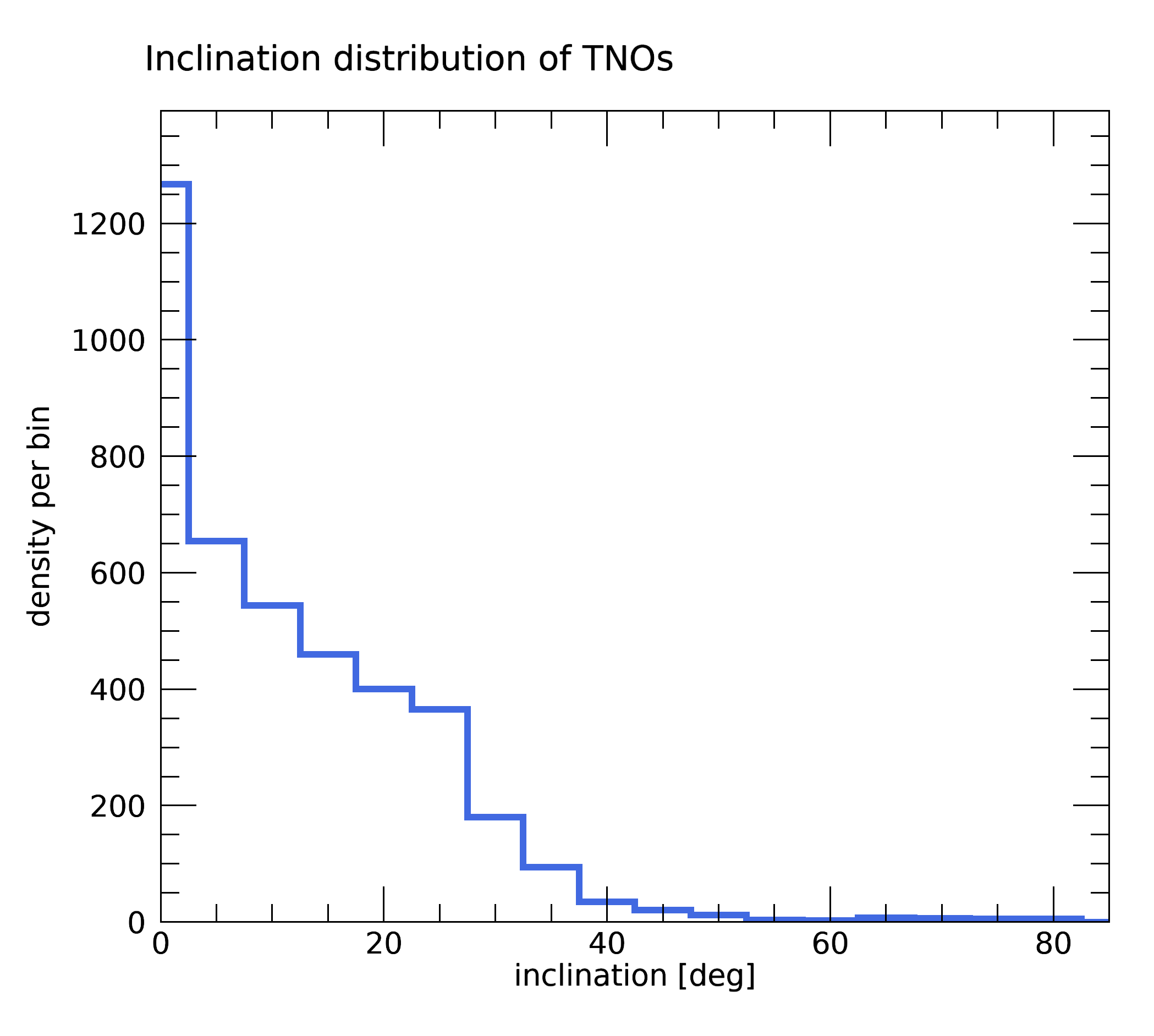}
\caption{\label{fig:ecc-inc-dist} 
The number distribution of the 4121 TNO of our sample, with respect to the eccentricities and inclinations.
}
\end{figure}

For an appropriate choice of the initial orbital elements of the test particles, in particular of $ e $ and $ I $, we investigated the distribution of the orbital elements of the real TNOs. Figure~\ref{fig:ecc-inc-dist} displays their number distribution with respect to their eccentricities and inclinations. Based on these distribution plots, for the region $34\leq a \leq 49$ AU closer to Neptune, we selected the eccentricities between $0\leq e \leq 0.6$, while in the more distant region $50\leq a \leq 80$ AU, the eccentricities were taken from a larger range of $0\leq e \leq 0.8$.

As for the selection of the orbital inclinations $ I $, we considered two different cases. In the first one, both the inclinations $I$ and the longitudes of ascending node $\Omega$ were set to $0^\circ$. In the second case, however, the inclinations were chosen randomly within the interval $0^\circ \leq I \leq 30^\circ$ (based on the second panel of Figure~\ref{fig:ecc-inc-dist}). The longitudes of ascending nodes were again set to $\Omega = 0^\circ$. The rest of the orbital elements, i.e., the argument of perihelion $\omega$ and the mean anomaly $M$ were randomly selected between $0^\circ$ and $360^\circ$. 

\begin{figure*}
\centering
\includegraphics[width=0.45\linewidth]{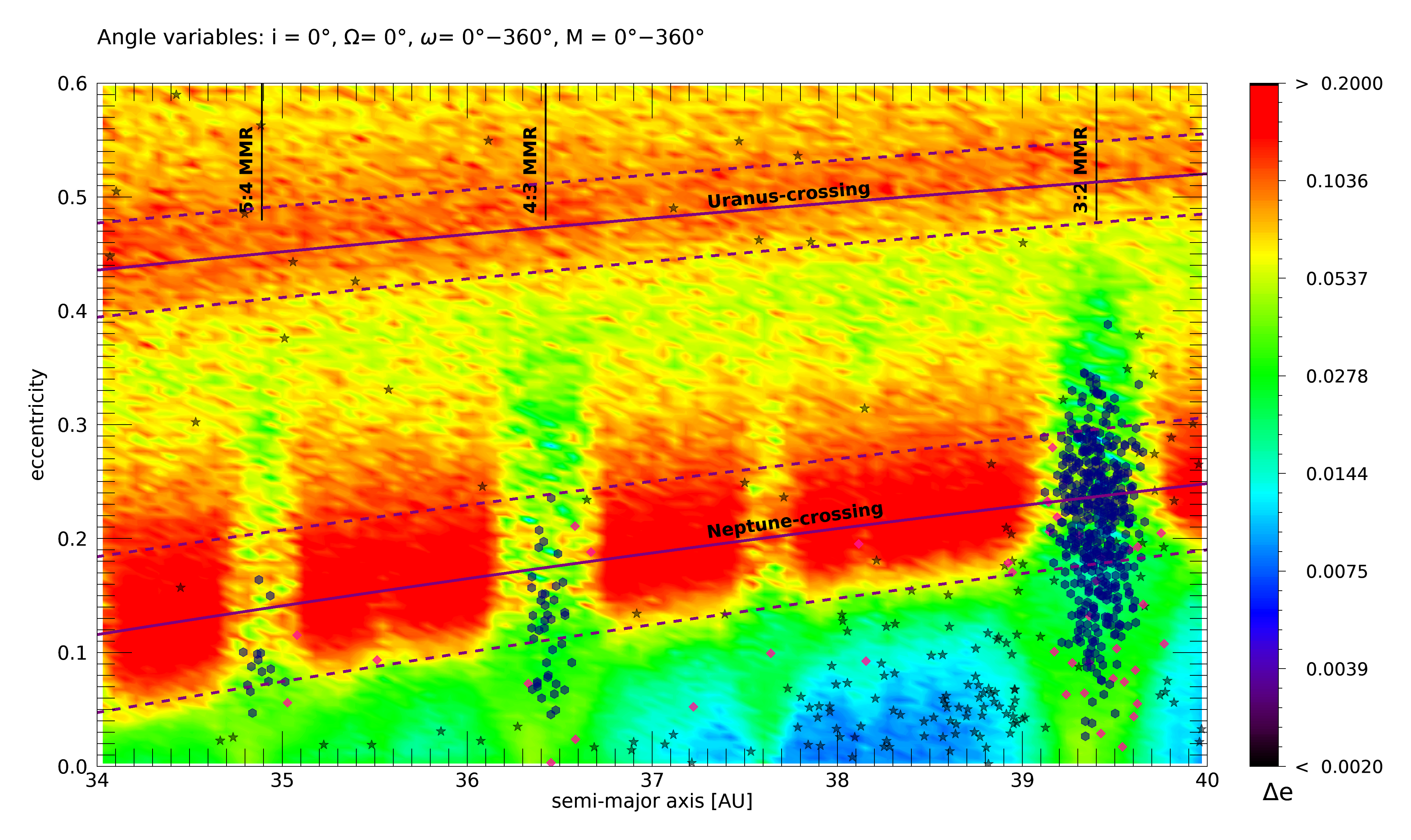}
\includegraphics[width=0.45\linewidth]{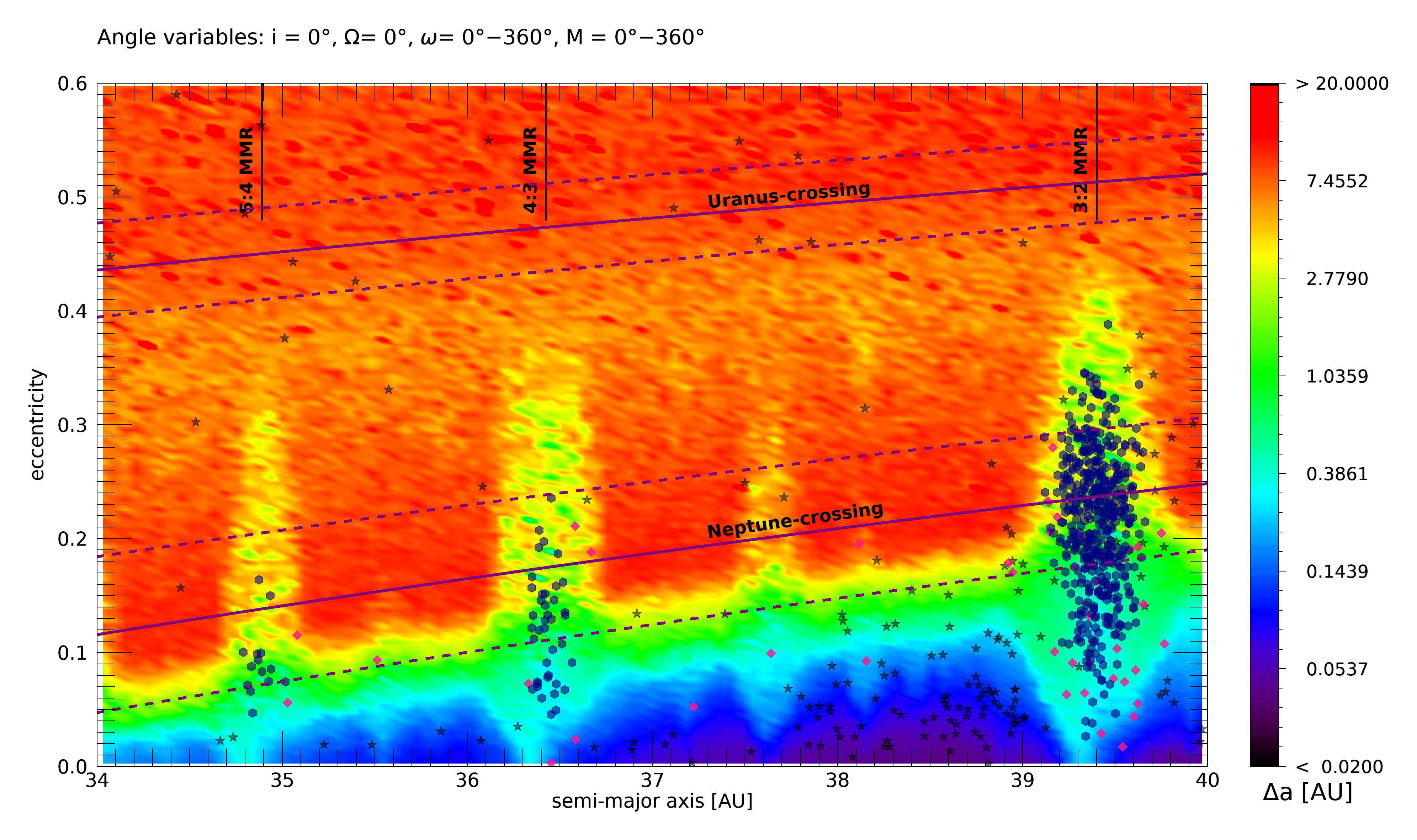}
\caption{\label{fig:dynmap-sma-34-40-i-0} 
Dynamical maps of the trans-Neptunian region between $34 \text{ AU} \leq a \leq 40 \text{ AU}$ and $0 \leq e \leq 0.6$. The inclinations and the longitudes of ascending node are kept fixed at $I=0^\circ$ and $\Omega = 0^\circ$. The arguments of perihelion $\omega$ and the mean anomalies $M$ are chosen randomly. Blue hexagons: initial $ (a, e) $ values of TNOs in long-term resonances; pink diamonds: initial $ (a, e) $ values of TNOs in short-term resonances; black (/white - in subsequent figures) stars: initial $ (a, e) $ values of non-resonant TNOs. 
Solid lines: $(a,e)$ values resulting in Neptune- or Uranus-crossing orbits. Dashed lines: distances of three Hill radii from the two giant planets. Vertical lines on the top of the panels: MMRs hosting more than 10 TNOs. Left panel: map of $\Delta e$. Right panel: map of $\Delta a$.
}
\end{figure*}

\begin{figure*}
\centering
\includegraphics[width=0.45\linewidth]{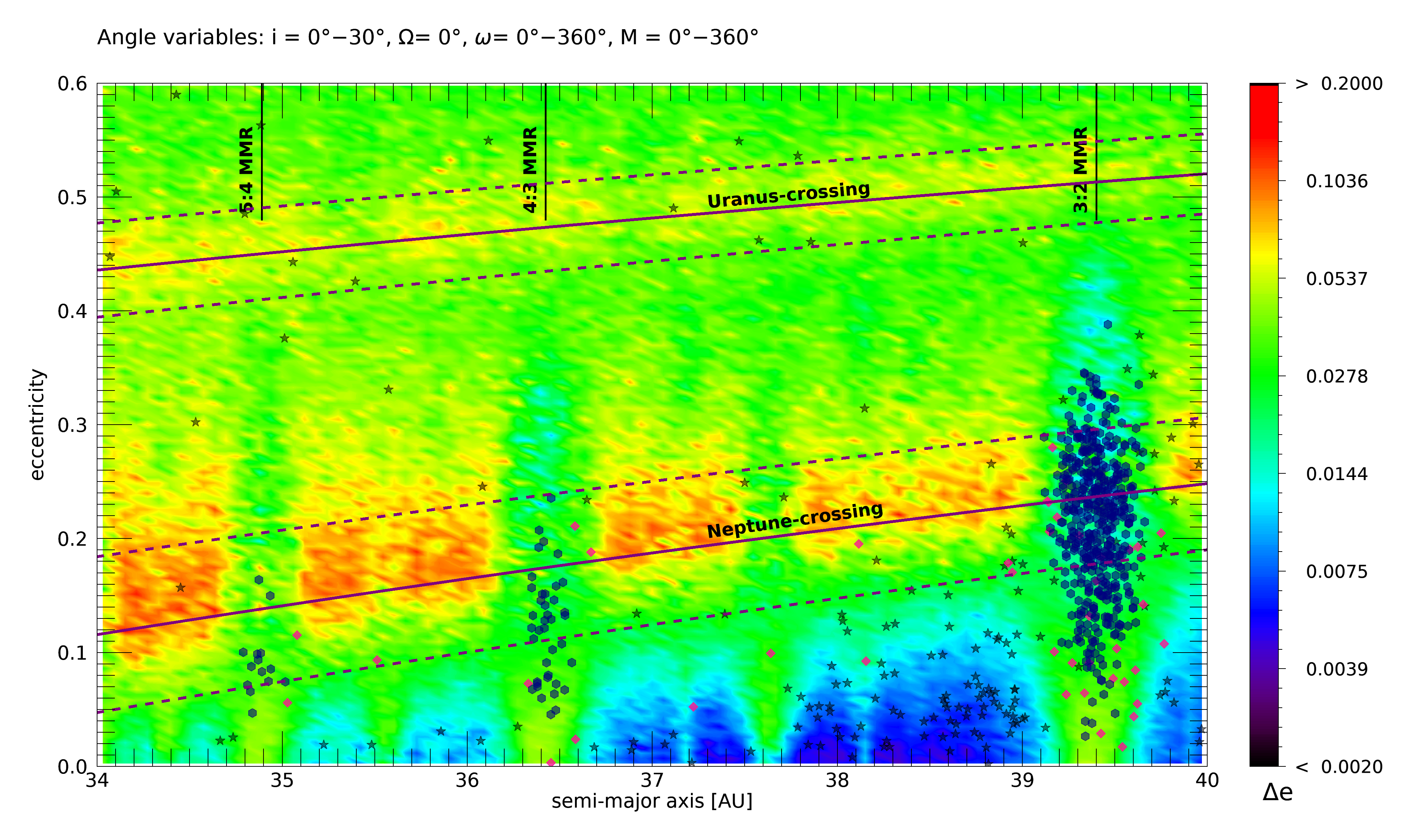}
\includegraphics[width=0.45\linewidth]{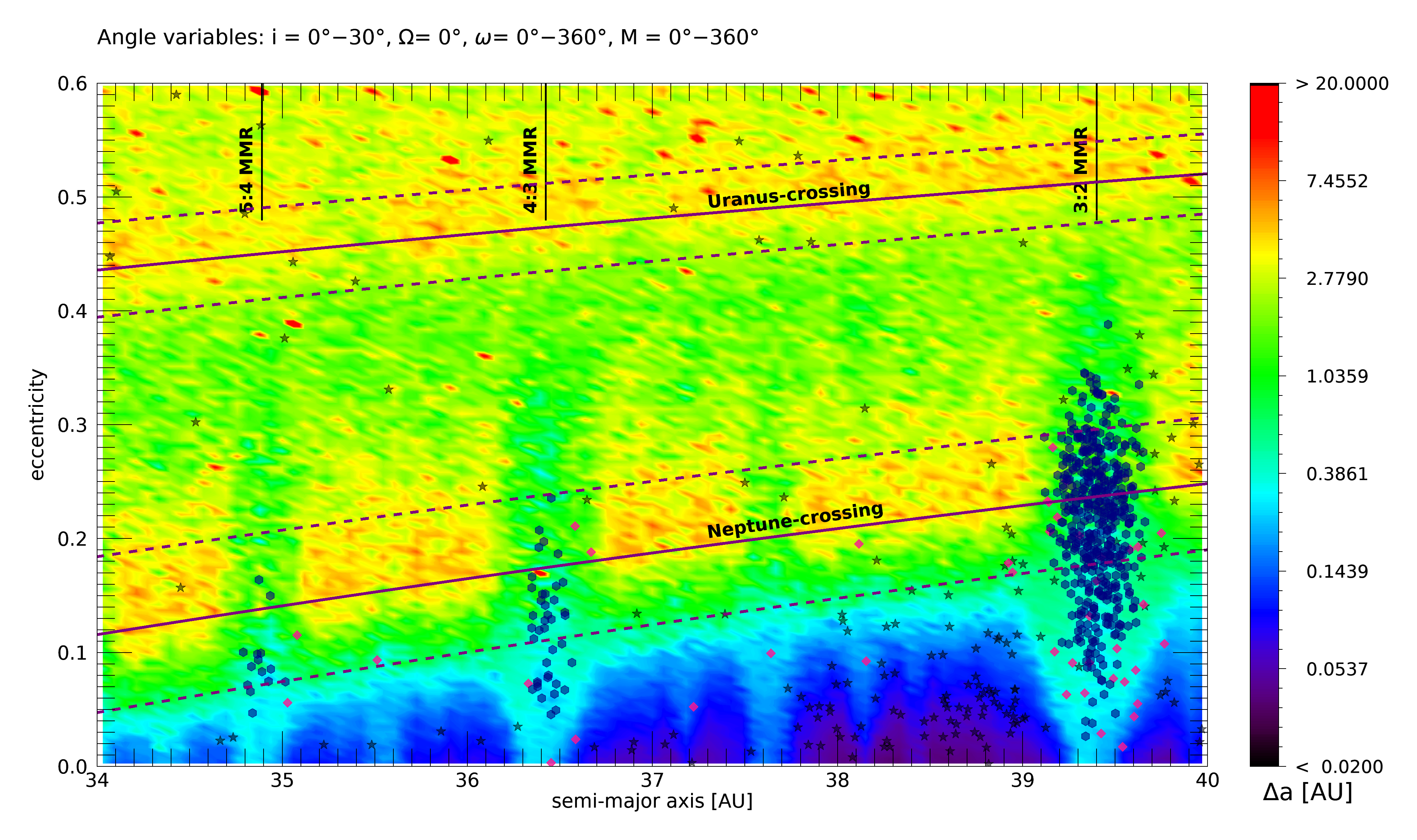}
\caption{\label{fig:dynmap-sma-34-40-i-30} 
The same as Figure~\ref{fig:dynmap-sma-34-40-i-0} except for the initial inclination of the test particles chosen randomly between $0^\circ \leq I \leq 30^\circ$.
}
\end{figure*}

In addition to the calculation of the maximum eccentricity variations $\Delta e$, we also calculated the maximum variation of the test particles' semi-major axis $ \Delta a $ (see again Section~\ref{sec:data_and_methods}), to get a more complete dynamical picture of the neighbourhood of the MMRs. The reason for the use of $\Delta a$ as an indicator of chaos is hinted by the third panels of Figures \ref{fig:3-2-MMR-Both-Pluto}-\ref{fig:13-9-MMR-2014QH563}. These plots depict the TNOs' excursion in the $(a,e)$ plane, and one observes significant changes not only in the $ e $ but also in the $ a $ values. Furthermore, by noting the colouring, too, of the panels, one can deduce the time spent in the different subdomains of the covered regions. The comparison of this latter property in the case of these 8 figures suggests that it is worth sentencing some special attention to the quantity $ \Delta a $, too.

After discussing the technical details with regard to the construction of the dynamical maps, we can now focus on the related results. The first two pairs of dynamical maps to be presented in Figures~\ref{fig:dynmap-sma-34-40-i-0} and \ref{fig:dynmap-sma-34-40-i-30}, show the region between $34 \mathrm{AU} \leq a \leq 40$ AU. Figure~\ref{fig:dynmap-sma-34-40-i-0} stands for the case $I=0^\circ$, whereas in Figure~\ref{fig:dynmap-sma-34-40-i-30} $ I $ is varied between $0^\circ \leq I \leq 30^\circ$. The first panel of each figure uses $\Delta e$ as a chaos indicator, while the second one portrays $\Delta a$. Apart from the test particles, i.e. the grid points, the real TNOs are also displayed in the figures. The blue hexagons denote objects with long-term librating critical arguments, while the pink diamonds mark TNOs with short-term librating critical arguments. For completeness, we also portray the non-resonant TNOs by black stars (in the subsequent figures the coloring is white). The more populated MMRs with more than 10 members are also highlighted by vertical lines. Furthermore, we display with solid lines those $(a,e)$ values that result in Neptune- or Uranus-crossing orbits. The dashed lines around them indicate the distances of three Hill radii to Neptune and Uranus. To make the fine structure of the dynamical maps prominent we use a logarithmic colour scale.
 
The observation of Figure \ref{fig:dynmap-sma-34-40-i-0} shows that both indicators outline the more stable domains in the vicinity of the resonances as well as the unstable regions arising from perturbations of close encounters with the giant planets. The general stabilizing role of the resonances is apparent as they penetrate into the chaotic domains. The curve-like unstable areas of three Hill radii of either Neptune or Uranus are intersected by the candle-like shapes of the MMRs where the resonant TNOs reside, too. As for the distinction of TNOs of short- and long-term resonances, the former ones are generally located near the edges of the resonances, although counterexamples can also be found. The detailed explanation of this finding will be investigated in our forthcoming publication.

\begin{figure*}
\centering
\includegraphics[width=0.45\linewidth]{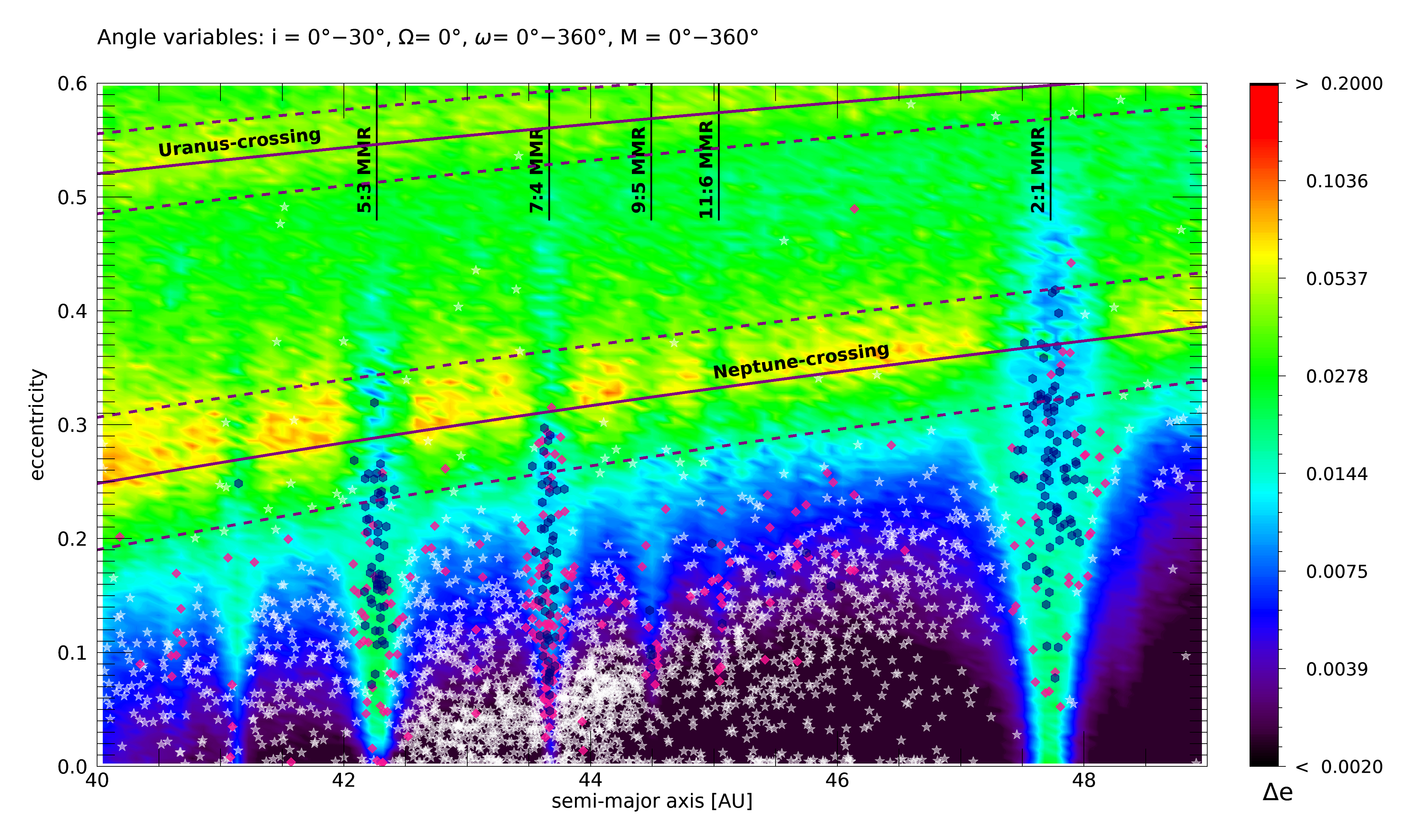}
\includegraphics[width=0.45\linewidth]{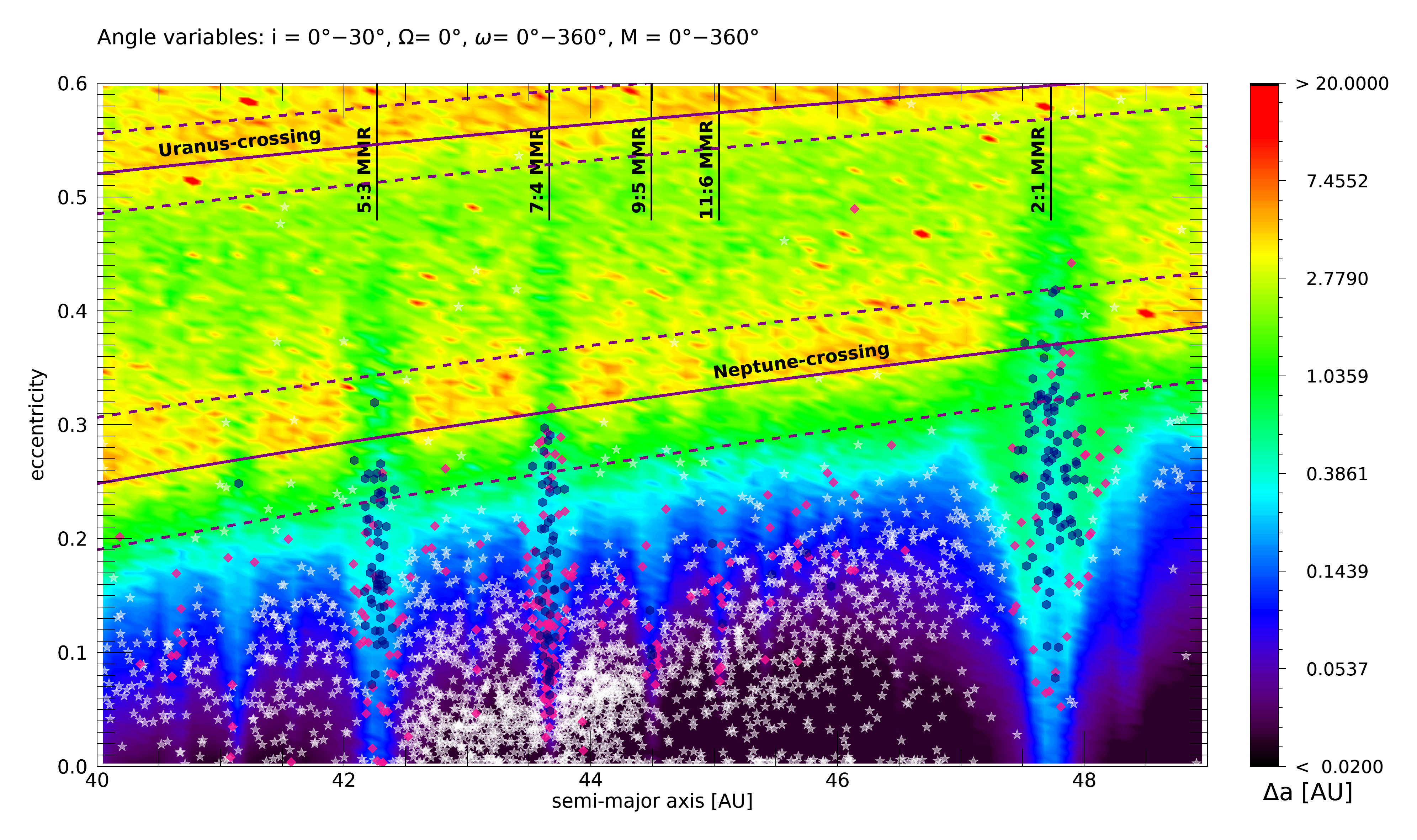}
\caption{\label{fig:dynmap-sma-40-49-i-30} 
The same as Figure~\ref{fig:dynmap-sma-34-40-i-30} except for the range in the semi-major axis: $40 \text{ AU} \leq a \leq 49 \text{ AU}$.
}
\end{figure*}

\begin{figure*}
\centering
\includegraphics[width=0.45\linewidth]{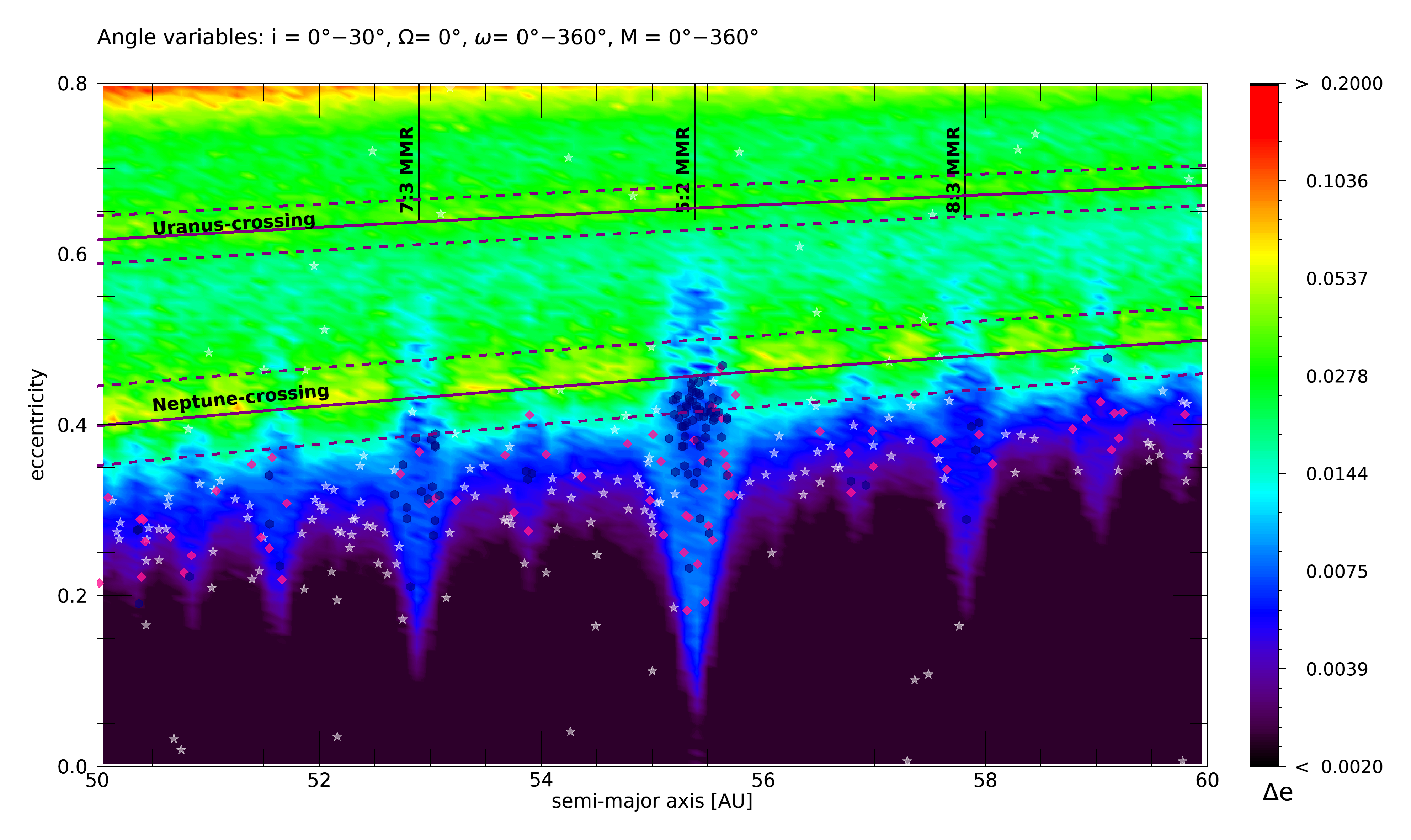}
\includegraphics[width=0.45\linewidth]{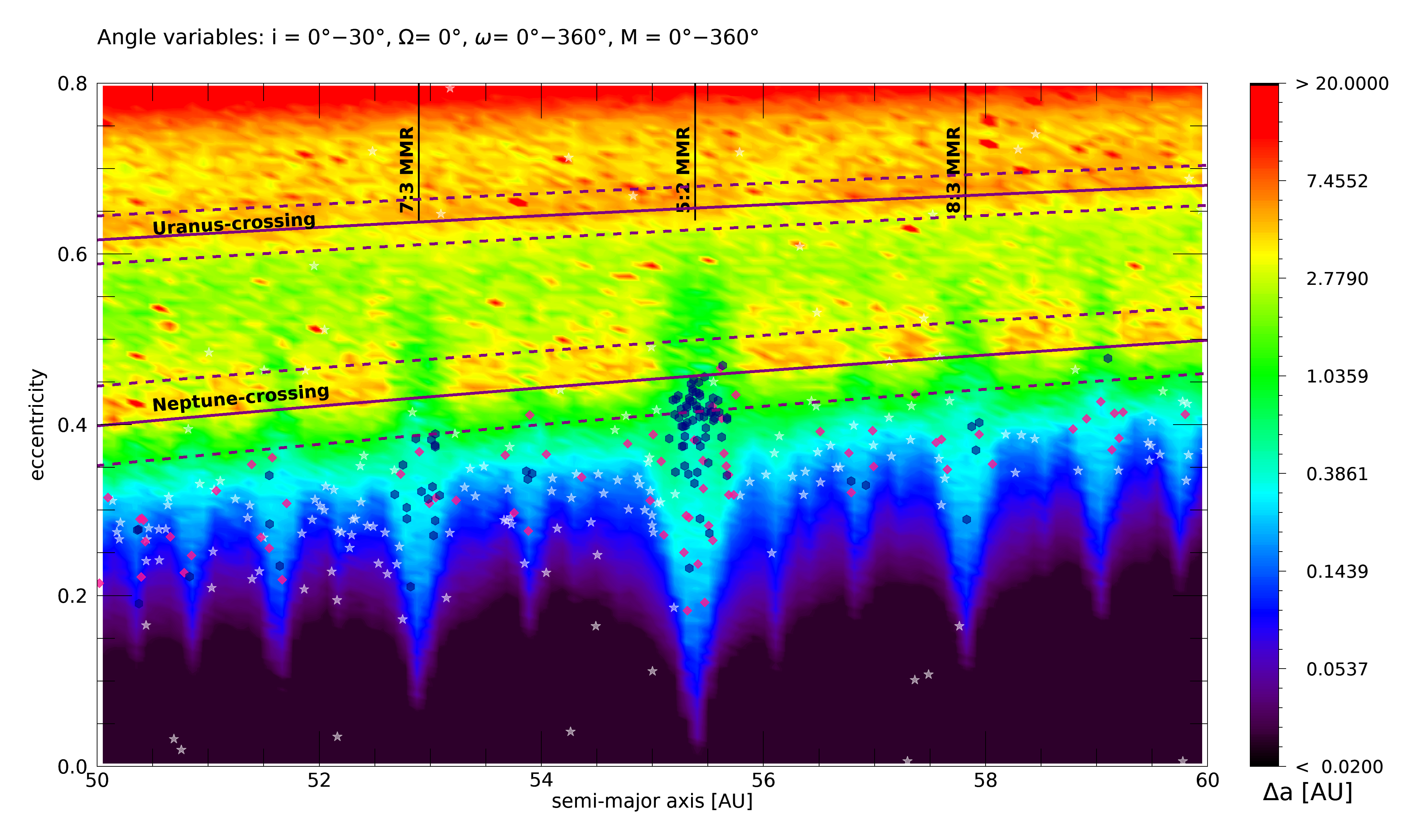}
\caption{\label{fig:dynmap-sma-50-60-i-30} 
The same as Figure~\ref{fig:dynmap-sma-34-40-i-30} except for the range in the semi-major axis: $50 \text{ AU} \leq a \leq 60 \text{ AU}$.
}
\end{figure*}

\begin{figure*}
\centering
\includegraphics[width=0.45\linewidth]{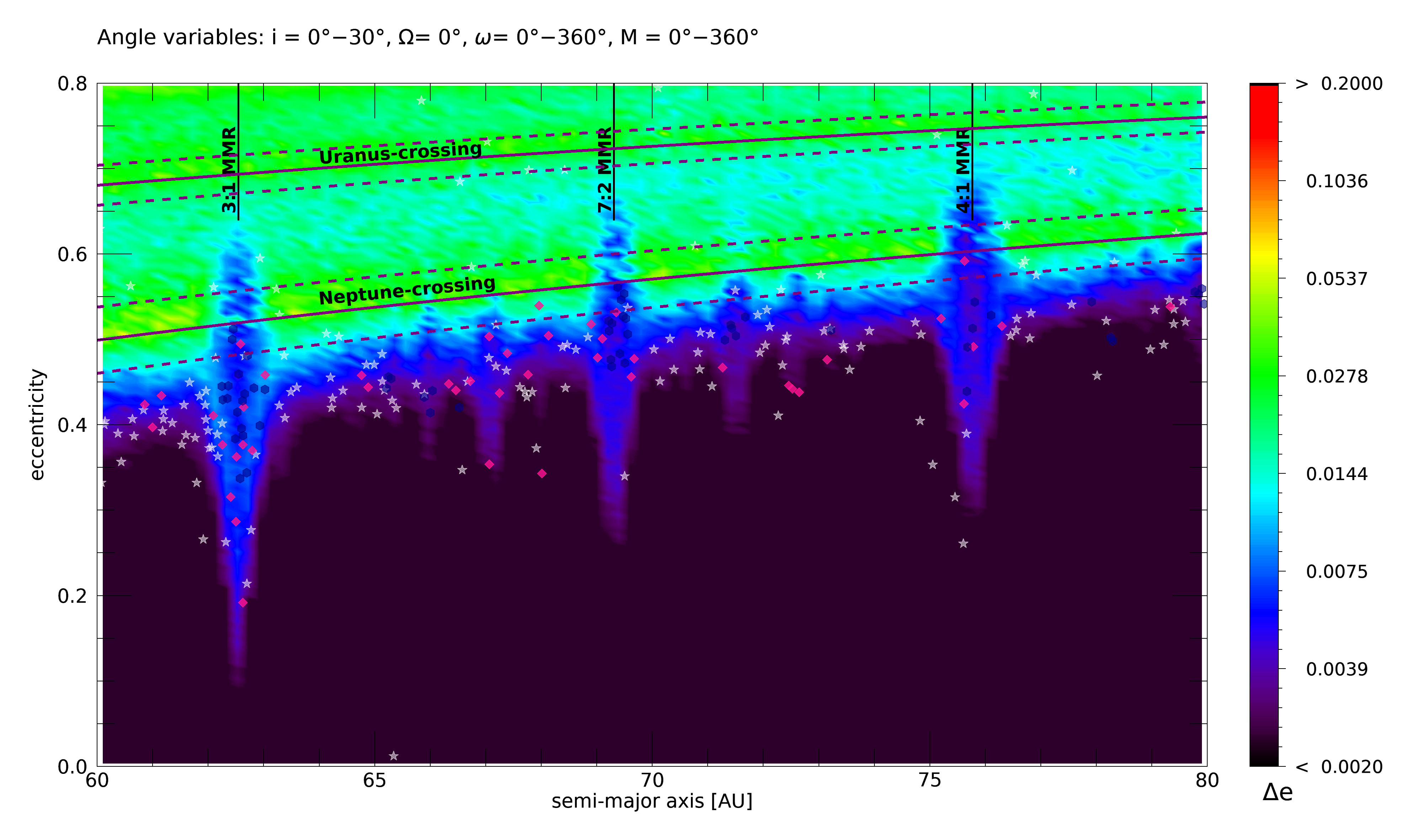}
\includegraphics[width=0.45\linewidth]{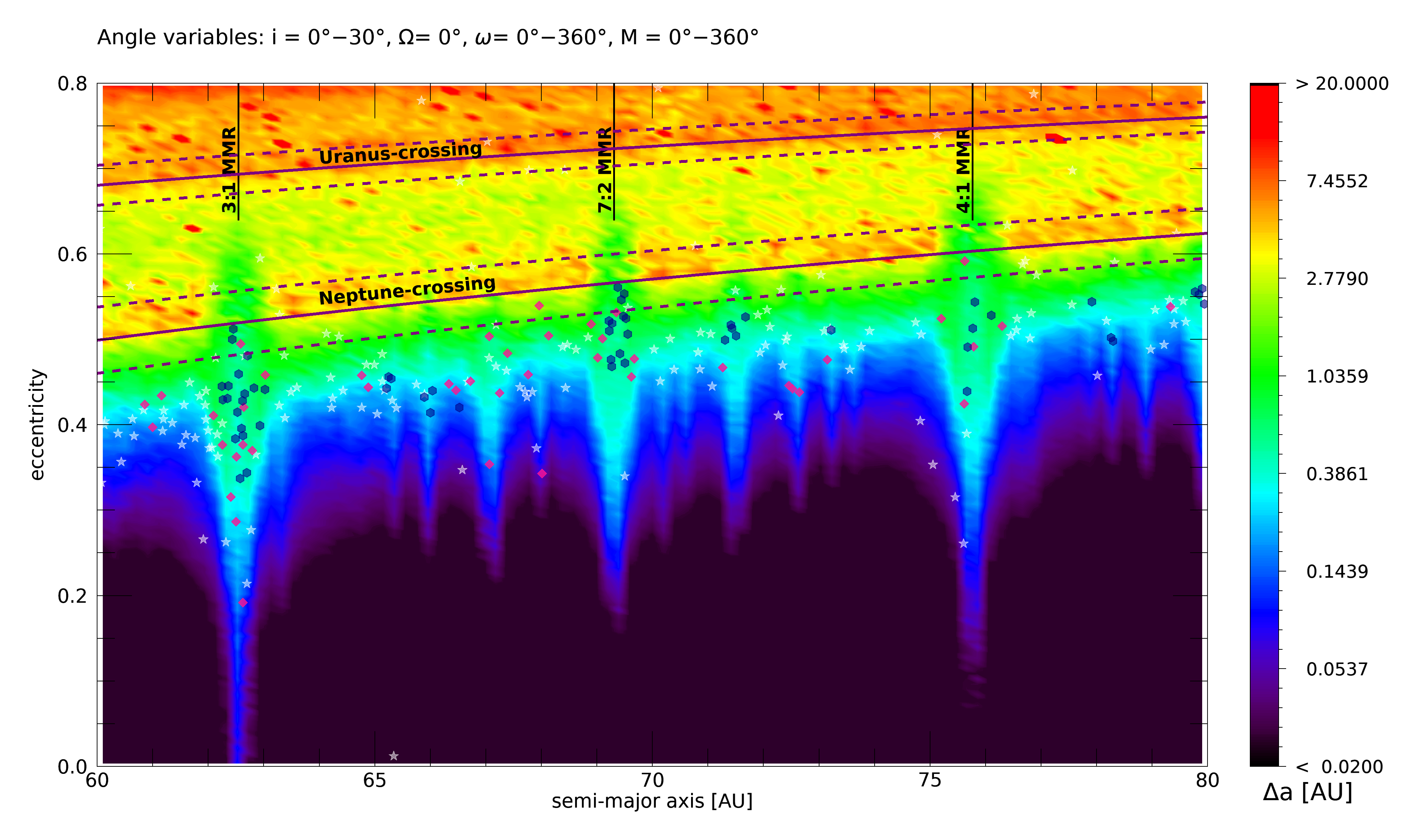}
\caption{\label{fig:dynmap-sma-60-80-i-30} 
The same as Figure~\ref{fig:dynmap-sma-34-40-i-30} except for the range in the semi-major axis: $60 \mathrm{AU} \leq a \leq 80 \mathrm{AU}$.
}
\end{figure*}

By comparing the dynamical maps of the indicators $\Delta e$ and $\Delta a$, one can conclude that the $\Delta a$ is "more sensitive". It reveals more of the fine structure of the phase space. One such interesting fine-structure phenomenon is that the V-shape of the MMRs is slightly asymmetric (see in particular the low-eccentricity parts of the figures where it is clearly discernible that the left-hand sides of the resonances are more elongated in the direction of $ e = 0 $). Apparently, both in the cases of the 4:3 and 3:2 MMRs, primarily short-term TNOs are situated in this area. This asymmetry holds, too, in the cases of the dynamical maps to be discussed below; therefore, we can state that the shape of the MMRs is more W- than V-like, with the right-hand side of the W being less-evolved at lower eccentricities. This asymmetric feature is not so markedly observable in the $\Delta e_\mathrm{max}$ maps.

Studying Figure~\ref{fig:dynmap-sma-34-40-i-0} in comparison with Figure~\ref{fig:dynmap-sma-34-40-i-30}, the importance of the choice of the angle variables, especially that of the inclination, is striking. (It should be noted here that to facilitate the comparison of the different figures, we used the same logarithmic colour scale for all dynamical maps.) If the initial value of the inclination of the test particles is not set to zero, the previously very unstable areas (red) are remarkably reduced and limited to the three-Hill-radii regions of the giant planets. In addition, the W-shaped bottom parts of the MMRs also become slightly wider.

Figures~\ref{fig:dynmap-sma-40-49-i-30}-\ref{fig:dynmap-sma-60-80-i-30} depict the dynamics of increasingly distant regions; the domains of 40-49 AU, 50-60 AU, and 60-80 AU, respectively. In the last two cases, the eccentricity range is also extended from the previous 0-0.6 to 0-0.8. The more detailed property of the $\Delta a$ maps is even more noticeable in these figures. New MMRs emerge at lower eccentricities that are otherwise difficult to discover in the $\Delta e$ maps. The resonance overlaps in the high-eccentricity region are nicely accentuated, too. Apparently, there exist several resonances in these remote regions; however, the detection of such faraway TNOs is still observationally difficult.

In conclusion, in this section we presented the dynamical maps of the $34 \mathrm{AU} \leq a \leq 80$ AU region, providing a complex picture of these outermost realms of the Solar System. In the given figures, we indicated the TNOs engaged in MMRs and made a distinction between those bodies that are characterized by a resonant angle librating during the entire time span of the integration, and those whose resonant angle exhibits only temporary libration. The maps clearly demonstrate the stabilizing role of the MMRs, even in the regions where close encounters with the giant planets would otherwise result in fatal perturbations, collisions, or expulsions. Yet some counterexamples of the above protective mechanism were also given, see e.g. the well-detailed case of the Trojan asteroid 2013\,TK$_{227}$ in Figure~\ref{fig:1-1-MMR-2013TK227} among others. We also saw that the non-resonant TNOs accumulate around but outside the MMRs. Only a few of them show up inside the resonant V-shapes. We stated before that the necessary (but not sufficient) condition for being in an MMR is to be in a mean-motion commensurability. That is, inside the mean-motion commensurabilities TNOs with non-librating critical arguments can also reside. Among the non-resonant TNOs those close to the edges of the MMRs deserve special attention for they can easily change between the resonant and non-resonant character by slow phase-space diffusion.

We note here that, by placing a real TNO in a dynamical map, the test particles in its neighbourhood can be considered as its clones. Hence by using these maps, one can obtain a more realistic picture of the dynamics of the real TNOs, too. We are aware of the fact that these dynamical maps do not substitute such dynamical investigations, in which thousands of clones or shadow particles are involved; however, they can be regarded as a lower resolution ``zeroth approach”.

\section{Summary and discussion\label{sec:summa}}

In this paper, we have extensively studied the dynamics of trans-Neptunian objects involved in mean-motion resonances with Neptune. In the analytic description of planetary motion, three types of MMRs appear, in explicit forms:
the pure eccentricity-type resonances include the longitudes of perihelion (of the asteroid and/or of Neptune) in the critical arguments of the disturbing function, while the inclination-type resonances contain only the ascending nodes of (either of) the bodies. The mixed-type resonances, not studied in the present work, incorporate both.

Although the first two types of the above resonances are described by different critical arguments (see Equations~\eqref{eq:crit_arg_e}-\eqref{eq:crit_arg_eprime} and \eqref{eq:crit_arg_inc}-\eqref{eq:crit_arg_incprime}), we found that inclination-type resonances usually do not appear alone but are accompanied by 
eccentricity-type ones, with a few exceptions. Furthermore, the inclination-type resonances accumulate around $e\approx 0.25$ and are not limited to large inclinations. That is, a significant amount of TNOs in inclination-type resonances orbit close to the invariable plane ($I<10^{\circ}$). This result seems to be a bit contradictory to the opinion that the inclination-type resonances manifest at large $I$s; however, as has been shown in the case of migrating giant planets, inclination-type MMRs can be excited when the bodies are captured in eccentricity-type MMRs during the migration \citep{2003ApJ...597..566T,2009MNRAS.400.1373L}. We note that among the resonant TNOs, no distinction has been made so far between the eccentricity- and inclination-type MMRs. Thus the presence of TNOs involved in both types of MMRs may favour the scenario that these bodies have been captured into the various MMRs during the outward migration of Neptune. In the other hand, the low values of the inclinations may indicate the presence of a possible damping of the inclination after Neptune's outward migration. 

Another remark to be made here is that both the eccentricity- and inclination-type MMRs can be further divided according to whether the coefficients of the critical arguments of the perturbing function contain only the eccentricity $ e' $ of Neptune, only the eccentricity $ e $ of the TNO, or the product of them. Similarly, among the mixed-type MMRs several combinations of $ e, e', I, $ and $ I' $ are allowed. In our work, we considered only the ''pure" types of the eccentricity- and inclination-type resonances (see Eqs. \eqref{eq:crit_arg_e}-\eqref{eq:crit_arg_incprime}). These additional splittings could also be taken into account.

It is also interesting to reveal the various secular effects working inside the MMRs. For instance, \cite{2022AJ....164...74L} carried out semi-analytic investigations of the von Zeipel--Lidov--Kozai (ZLK) effect. Independently of one another, \cite{1910AN....183..345V}, \cite{1962P&SS....9..719L} and \cite{1962AJ.....67..591K} studied the possible connections between the oscillation of eccentricities and inclinations during the long-term evolution of small bodies. Furthermore, \cite{2020tnss.book...61K} claimed that the ZLK effect might be important to understand the distribution of the TNOs' orbital elements. The impact of the ZLK effect on the TNOs has also been studied by \cite{2020CeMDA.132...12S}. In this work, a two-degrees-of-freedom semi-secular model has been adopted by averaging the short-period terms. In an analytical treatment of this model, \cite{2022AJ....164...74L} applied a one-degree-of-freedom integrable model using the method of adiabatic invariants. The resulting phase portraits gave a good prediction of the long-term evolution of eccentricity, inclination, and longitude of perihelion.

Apart from distinguishing between the different types of MMRs, the further refinement of the dynamical behaviour of the TNOs was made by monitoring the time evolution of the critical arguments $(\Theta^{(e)}_{1},\,\Theta^{(e')}_{2})$ or $(\Theta^{(I)}_{1},\,\Theta^{(I')}_{2}).$ We designate as long-term libration of the critical argument when it oscillates around a mean value throughout the entire time span of the integration, whereas a temporary oscillation - interrupted by shorter/longer intervals of circulation - is called short-term libration. Interestingly, these two kinds of libration are clearly separated in the case of the inclination-type MMRs: short-term libration appears at high inclinations ($I>10^{\circ}$), while long-term types manifest at lower values of $ I $. A general observation, with regard to the distinction between the short- and long-term types of the libration, is that the long-term librating TNOs gather in the inner parts of the mean-motion commensurabilities, while their short-term librating counterparts assemble close to the boundaries from where they can more easily escape or drift to another MMR.

As for the diffusion of the TNOs, \cite{2021AJ....162..164M} have investigated the dynamics and stability of the 34 largest objects in the trans-Neptunian region, referred to as Dwarf Planets (DPs). They took into account the gravitational interactions not only in between the giant planets but also in between the DPs. By considering 200 clones (called 'realisations' in their paper) for each DP, their long-term numerical integrations resulted in 17 stable, 11 unstable, and 6 resonant objects. They found that two objects, namely the TNOs 2010\,RF$_{43}$ and 2002\,UX$_{25}$, leave statistically the trans-Neptunian region and presumably the entire Solar System within $10^9$ years, it is however unlikely that they drift toward the inner Solar System.
Yet further investigations of the chaotic diffusion in the trans-Neptunian region could still be addressed.

All in all, our comprehensive analysis of 4121 individual TNOs (accompanied by dynamical maps of test particles) outlines some fundamental dynamical behaviours of the trans-Neptunian space. The future large-scale surveys (such as the Vera Rubin Observatory LSST project) are going to provide a remarkably larger sample of TNOs and thus the refined statistics can be used to improve our knowledge on the origins and fate of minor bodies in the outer Solar System.

\begin{acknowledgments}
We acknowledge the computational resources of the GPU Laboratory of the Wigner Research Centre for Physics. Furthermore, E. K. acknowledges the support of the ÚNKP-21-3 and ÚNKP-22-3 New National Excellence Programs of the Ministry for Innovation and Technology from the source of the National Research, Development, and Innovation Fund. C. K. has been supported by the National Research, Development, and Innovation Office (NKFIH), Hungary, through the grant K-138962.

\end{acknowledgments}

\bibliography{main}{}
\bibliographystyle{aasjournal}

\end{document}